\documentclass[a4paper,11pt]{article}
\pdfoutput=1 

\usepackage{jheppub} 
\usepackage[T1]{fontenc} 
\usepackage{amssymb}
\usepackage{amsmath}
\usepackage{slashed}
\usepackage{graphicx}

\def\ba{\begin{eqnarray}}
\def\ea{\end{eqnarray}}

\def\be{\begin{equation}}
\def\ee{\end{equation}}

\def\sumint{\hbox{$\sum$}\!\!\!\!\!\!\!\,{\int\!}}

\def\half{{\textstyle \frac 12}}

\def\symff{\text{SYM}_{4,4}}
\def\symod{\text{SYM}_{1,{\cal D}}}
\def\symot{\text{SYM}_{1,10}}

\newcommand{\la}{\label}

\begin{document}

\title{${\cal N}=4$ supersymmetric Yang-Mills thermodynamics to order $\lambda^2$}

\author[a,b]{Qianqian Du,}
\author[b]{Michael Strickland,}
\author[b]{and Ubaid Tantary}
\affiliation[a]{Institute of Particle Physics and Key Laboratory of Quark and Lepton Physics (MOS), Central China Normal University, Wuhan, 430079, China}
\affiliation[b]{Department of Physics, Kent State University, Kent, OH 44242, United States}
\emailAdd{duqianqianstudent@mails.ccnu.edu.cn}
\emailAdd{mstrick6@kent.edu}
\emailAdd{utantary@kent.edu}

\abstract{
We calculate the resummed perturbative free energy of ${\cal N}=4$ supersymmetric Yang-Mills in four spacetime dimensions ($\symff$) through second order in the 't Hooft coupling $\lambda$ at finite temperature and zero chemical potential. Our final result is ultraviolet finite and all infrared divergences generated at three-loop level are canceled by summing over $\symff$ ring diagrams. Non-analytic terms at ${\cal O}({\lambda}^{3/2}) $ and $ {\cal O}({\lambda}^2 \log\lambda )$ are generated by dressing the $A_0$ and scalar propagators. The gauge-field Debye mass $m_D$ and the scalar thermal mass $M$ are determined from their corresponding finite-temperature self-energies. Based on this, we obtain the three-loop thermodynamic functions of $\symff$ to ${\cal O}(\lambda^2)$.  We compare our final result with prior results obtained in the weak- and strong-coupling limits and construct a generalized Pad\'{e} approximant that interpolates between the weak-coupling result and the large-$N_c$ strong-coupling result. Our results suggest that the ${\cal O}(\lambda^2)$ weak-coupling result for the scaled entropy density is a quantitatively reliable approximation to the scaled entropy density for $0 \leq \lambda \lesssim 2$.
}

\date{\today}


\maketitle

\newpage

\section{Introduction}
\label{sect:intro}

The perturbative expansion of the free energy of $\mathcal{N}=4$ supersymmetric Yang-Mills in four dimensions ($\symff$) with $N_c$ colors and gauge coupling $g$ can be written in the form
\ba\la{highT}
\lim_{\lambda \rightarrow 0} {\cal F} \sim T^4 \big[ a_0 + a_2 \lambda + a_3 \lambda^{3/2} + \big( a_4+ a_4^{\prime} \log  \lambda \big) \lambda^2 + \mathcal{O}(\lambda^{5/2}) \big]  \, ,
\ea
where $\lambda=g^2 N_c$ is the 't Hooft coupling, which does not run and is independent of the temperature.\footnote{The notation SYM$_{x,y}$ indicates ${\cal N}=x$ supersymmetric Yang-Mills theory in $y$ spacetime dimensions.}   This expansion is identical in form to the perturbative expansion of the QCD free energy \cite{Arnold:1994ps,Arnold:1994eb,Zhai:1995ac,Braaten:1995jr,Kajantie:2002wa}.  The leading term in this expression is the free energy of an ideal $\symff$ plasma and the ${\cal O}(\lambda)$ correction can be obtained by computing two-loop Feynman diagrams.   Naively, the next contribution is ${\cal O}(\lambda^2)$ and comes from three-loop contributions.  However, a problem emerges because one finds uncanceled infrared divergences at the three-loop level if one uses  bare propagators.   The same issue occurs in QCD and, in this case, the infrared divergences can be eliminated by summing over the so-called ring diagrams~\cite{Arnold:1994ps,Arnold:1994eb,Zhai:1995ac,Braaten:1995jr}.  The solution is the same in $\symff$, with the key difference between QCD and $\symff$ being the number and types of degrees of freedom, since the $\symff$ theory has six scalar fields and four Majorana fermions. In order to cancel the three-loop infrared divergences, similar to the case of QCD, one can reorganize the perturbative calculation to take into account the soft thermal masses of the gluon and scalar fields \cite{Gross:1980br,kapusta1993finite,Andersen:2004fp}.  In QCD, through ${\cal O}(\lambda^{5/2})$, these soft mass scales result in non-analytic contributions to the free energy at orders $\lambda^{3/2}$, $\lambda^2 \log \lambda$, and $\lambda^{5/2}$ \cite{Arnold:1994ps,Arnold:1994eb,Zhai:1995ac,Braaten:1995jr,Ghiglieri:2020dpq} and an analytic contribution at order $\lambda^2$.

In the weak-coupling limit, the free energy of $\symff$ has been calculated through order $\lambda^{3/2}$, with the result being~\cite{Fotopoulos:1998es,Kim:1999sg,VazquezMozo:1999ic,Nieto:1999kc}
\ba\la{weaexp}
\frac{\cal F}{{\cal F}_{\textrm{ideal}}}= \frac{\cal S}{{\cal S}_{\textrm{ideal}}} =1-\frac{3}{2\pi^2}\lambda+\frac{3+\sqrt{2}}{\pi^3}\lambda^{3/2} + \cdots   \, ,
\ea
where  $\mathcal{F}_{\textrm{ideal}} = - d_A \pi^2T^4/6$ is the ideal or Stefan-Boltzmann limit of the free energy and $\mathcal{S}_{\textrm{ideal}} = 2 d_A \pi^2T^3/3$, with $d_A = N_c^2 -1$ being the dimension of the adjoint representation. The first three terms in (\ref{weaexp}) map to $a_0=1$, $a_2=-\frac{3}{2\pi^2}$ and $a_3=\frac{3+\sqrt{2}}{\pi^3}$ in eq.~(\ref{highT}). The aim of our work is to compute the coefficients $a_4$ and $a_4^{\prime}$ in eq.~(\ref{highT}).

For this purpose, we will make use of the \emph{regularization by dimensional reduction} (RDR) method to (1) simplify some aspects of the calculation using dimensional reduction and (2) regulate any intermediate divergent momentum-space integrals encountered during the calculation \cite{Siegel:1979wq}. The dimensional reduction method was first discussed by Brink, Schwarz, and Scherk for supersymmetric Yang-Mills theories in 2, 4, 6, and 10 dimensions, and was applied to obtain various Yang-Mills theories with extended supersymmetry in 2 and 4 dimensions \cite{Brink:1976bc}. Using this method one can show that $\symff$ can be obtained by dimensional reduction of 10-dimensional $\mathcal{N}=1$ supersymmetric Yang-Mills theory ($\symot$).   We will make use of this fact to simplify the calculation of the three-loop $\symff$ diagrams by instead computing massless three-loop diagrams in $\symot$ followed by dimensional reduction to $\symff$ in the RDR scheme.   We note that there are inconsistencies in Siegel's RDR scheme which have been pointed out by several authors including Siegel himself \cite{Siegel:1980qs,Avdeev:1982xy,Stockinger:2005gx,Stockinger:2006kk}.  However, these inconsistencies only manifest themselves at high-loop order or under application of Fierz reordering identities and can be pushed to higher orders by making use of the superfield formalism \cite{Avdeev:1982xy}.  As discussed in Ref.~\cite{Avdeev:1982xy}, the breaking of supersymmetry in Siegel's RDR scheme first appears at 3-loop order in the ${\rm SYM}_{1,10} $ beta function and at higher-loop orders for other quantities such as the propagator of the vector supermultiplet.  Since a three-loop graph contributing to the running coupling maps to a four-loop vacuum contribution, the three-loop calculation of the thermodynamic potential presented herein should be free from such ambiguities.

In addition to the calculational simplification provided by dimensional reduction, the RDR scheme also provides a regularization method that manifestly preserves supersymmetry.  This method was first introduced by Siegel~\cite{Siegel:1979wq} and implements a modified form of dimensional regularization \cite{THOOFT1972189,Ashmore:1972uj,Bollini:1972ui} which manifestly preserves gauge invariance, unitarity, and  supersymmetry by making use of dimensional reduction~\cite{Brink:1976bc,Gliozzi:1976qd}.  The application of the RDR method to $\mathcal{N}=1, 2,$ and 4 SYM was considered by Avdeev and Vladimirov \cite{Avdeev:1982xy}, and the authors presented two equivalent methods for the evaluation of SYM Feynman diagrams based on dimensional reduction.  Since Siegel's original paper, other authors have considered variations of RDR in which the size of the representations can depend on the number of spatial dimensions, however, in these variants one must introduce additional degrees of freedom (so-called $\epsilon$-scalars) in order to preserve supersymmetry \cite{CAPPER1980479}.  This gives equivalent results to the original RDR scheme, but complicates the calculation since additional degrees of freedom and Feynman rules must be implemented.  For this reason, we make use of Siegel's original RDR scheme.  Put simply, in Siegel's RDR scheme, one maintains supersymmetry by keeping the size of the bosonic, fermionic, and scalar representations fixed when the number of spatial dimensions is changed.  As a result of this prescription, the cancellations between the bosonic and fermionic degrees of freedom necessary to maintain supersymmetry are automatically preserved.

In the resulting RDR scheme, when computing vacuum contributions to the free energy, one fixes the dimension of the field representations to be integer valued and takes all momentum to be $4-2\epsilon$ dimensional vectors, where $\epsilon$ is an infinitesimal.  Although the use of dimensional reduction results in a dramatic simplification in the computation of vacuum contributions since it is more straightforward to calculate higher-loop diagrams in $\symot$ than in $\symff$, one limitation of using the dimensional reduction is that, although there is exact equivalence between the results of vacuum graphs computed using dimensional reduction of 10-dimensional SYM$_1$ theory and the results of vacuum graphs computed directly in $\symff$, it is not possible to treat soft contributions to the free energy in the same manner.  For this purpose, one must introduce dressed gauge and scalar propagators directly in $\symff$ in order to resum the supersymmetric ring diagrams and eliminate infrared divergences generated at three-loop order.  The dimensional reduction method was first used for the computation of two-loop
$\symff$ thermodynamics in ref.~\cite{VazquezMozo:1999ic} where it was shown that, with this method, one can compute the hard contributions using dimensionally reduced $\symot$ and the soft contributions directly in $\symff$.  We will adopt the same strategy:  (1) use dimensional reduction applied to $\symot$ to compute the massless three-loop vacuum graphs and (2) compute the necessary soft contributions and counterterms directly in $\symff$ by dressing the gauge and scalar propagators.

Finally, we note that in the opposite limit of strong coupling, the behavior of the $\symff$ free energy has been computed using the anti-de Sitter space/CFT (AdS/CFT) correspondence \cite{Maldacena:1997re}.  In the large-$N_c$ limit one has~\cite{Gubser:1998nz}
\ba\la{stro}
\frac{\cal F}{{\cal F}_{\textrm{ideal}}}= \frac{\cal S}{{\cal S}_{\textrm{ideal}}} =\frac{3}{4}\bigg[1+\frac{15}{8}\zeta(3)\lambda^{-3/2} + {\cal O}(\lambda^{-3}) \bigg] ,
\label{eq:sclimit}
\ea
where, in the large-$N_c$ limit, it is expected that only powers of $\lambda^{-3/2}$ appear in the strong-coupling expansion.\footnote{If one considers sub-leading corrections in $1/N_c$ in the strong-coupling limit, the entropy density can contain additional fractional powers of $\lambda$ \cite{Myers:2008yi}.  It is also possible that there could be logarithms of the 't Hooft coupling induced by massless gravity modes in this case.}  In the results section we use this information to constrain the coefficients appearing in a large-$N_c$ generalized Pad\'{e} approximant constructed from terms of the form $\lambda^{n/2}$ and $\lambda^{n/2} \log\lambda$ with $n$ being a positive integer.  We demonstrate that the resulting generalized Pad\'{e} is free from singularities at all $\lambda$, reproduces the weak-coupling expansion through ${\cal O}(\lambda^2,\lambda^2 \log\lambda)$, and reproduces the strong-coupling expansion to all known orders, with no terms containing $\log\lambda$ through ${\cal O}(\lambda^{-5/2} \log\lambda)$.

The structure of our paper is as follows.  We begin with a brief summary of $\symot$ and its relation to $\symff$ in the RDR scheme. In sec.~\ref{N4_theory}, we recall the basics of $\symff$ and introduce the resummations necessary to obtain the three-loop corrections to the free energy. In sec.~\ref{potential}, we list all contributions to the free energy up to three-loop order and describe the key steps necessary to obtain our final results.  In this section we also present our final results at each loop order truncated at $\mathcal{O}(\epsilon^0)$.  In sec.~\ref{thermodynamics} we present our final result for the ${\cal O}(\lambda^2)$ and ${\cal O}(\lambda^2 \log\lambda)$ coefficients, present comparisons with past results, and compare to a generalized Pad\'{e} approximant constructed using the new constraints provided in the weak-coupling limit. Finally, in sec.~\ref{conclusions} we present our conclusions and an outlook for the future.

\subsection*{Notation and conventions}

We use lower-case letters for Minkowski space four-vectors, e.g. $p$, and upper-case letters for Euclidean space four-vectors, e.g. $P$.  We use the mostly minus convention for the metric.

\section{Dimensional reduction of $\symod$ in the RDR scheme}
\la{N1_theory}

The action of $\mathcal{N}=1$ supersymmetric Yang-Mills in ${\cal D}$ dimensions ($\symod$) can be written in Minkowski space as \cite{VazquezMozo:1999ic,Brink:1976bc}
\ba\la{actN1}
S_{\textrm{SYM}_{1,{\cal D}}} &=& \int d^\mathcal{D} x \,  \textrm{Tr}\bigg[{-}\frac{1}{2}G_{MN}^2+2i\bar{\psi} \Gamma^M D_M \psi \bigg] ,
\ea
where $M, N=0, \cdots, \mathcal{D}-1 $, and the field strength tensor is $G_{MN}=\partial_M A_N-\partial_N A_M-ig[A_M,A_N]$, and $D_M=\partial_M-i g[A_M,\cdot]$ is the covariant derivative in the adjoint representation of $SU(N_c)$.\footnote{In practice, the gauge group can be any semi-simple Lie group. Since our target theory is $\symff$ with the $SU(N_c)$ group, here we also use the $SU(N_c)$ group.} The definition of the gauge field is the same as QCD and $A_M$ can be expanded as \mbox{$A_M=A_M^a t^a$}, with real coefficients $A_M^a$, and Hermitian color generators $t^a$ in the fundamental representation that satisfy
\ba\la{A}
 [t^a,t^b]=i f_{abc}t^c   \quad \textrm{and} \quad \textrm{Tr}(t^a t^b)=\frac{1}{2}\delta^{ab} \, ,
\ea
where $a,b=1, \cdots ,N_c^2-1$.  The structure constants $f_{abc}$ are real and completely antisymmetric. The metric tensor is $g_{\scriptscriptstyle MN} =(1,-1,\cdots -1)$ and the $\Gamma$-matrices satisfy
\ba\la{gamma}
 \{ \Gamma_M, \Gamma_N \}=2 g_{\scriptscriptstyle MN} I_n \quad \textrm{with}\quad n=2^{\frac{\cal D}{2}}\, ,
\ea
with $I_n$ being an identity matrix of dimension $n$.

SYM theories with a different number of supercharges $\#_{SC}$ can be obtained by choosing the appropriate value of $\mathcal{D}$ for which the number of supercharges is maximal
\ba
\#_{SC}= 16 \rightarrow \mathcal{D}_{\textrm{max}} =10 \, ,  \quad\quad   \#_{SC} = 8 \rightarrow \mathcal{D}_{\textrm{max}} =6 \, ,  \quad\quad   \#_{SC} = 4 \rightarrow \mathcal{D}_{\textrm{max}} =4 \, .
\ea
In order to maintain supersymmetry, the number of bosonic and fermionic degrees of freedom must be equal, which implies that the fermions in $\mathcal{D}_{\textrm{max}} =10$ must satisfy the Majorana-Weyl condition, while they satisfy the Weyl condition in $\mathcal{D}_{\textrm{max}} =6$ and \mbox{$\mathcal{D}_{\textrm{max}} =4$} \cite{VazquezMozo:1999ic}.  Because of these constraints, in general, the number of degrees of freedom is equal to $\mathcal{D}_{\textrm{max}}-2$. In this way, one learns that the $n$ in (\ref{gamma}) should be the dimension of the spinors of the maximal SYM theory, and equal to $\mathcal{D}_{\textrm{max}}-2$, which implies $\textrm{Tr} \, I_n =\mathcal{D}_{\textrm{max}}-2$.

To quantize the theory, gauge-fixing and ghost terms should be added to the Lagrangian density.  In general covariant gauge they are the form
\ba\la{gfn1}
\mathcal{L}_{\textrm{gf}}^{\textrm{SYM}_{1,{\cal D}}}&=& -\frac{1}{\xi}\textrm{Tr}\big[(\partial^M A_M)^2 \big],  \nonumber \\
\mathcal{L}_{\textrm{gh}}^{\textrm{SYM}_{1,{\cal D}}} &=&-2\textrm{Tr}\big[\bar{\eta}\,\partial^M \! D_M \eta \big],
\ea
with $\xi$ being the gauge parameter.

In general, we will be interested in SYM theories with $\#_{SC}$ supercharges in dimensions $D\leq \mathcal{D}_{\textrm{max}}$, where $D$ is an integer. Any $D\leq \mathcal{D}_{\textrm{max}}$ dimensional SYM theory can be obtained by dimensional reduction from the corresponding maximal $\mathcal{N}=1$ SYM theory in $\mathcal{D}= \mathcal{D}_{\textrm{max}}$~\cite{Brink:1976bc}. To connect these different theories in a manifestly supersymmetric manner one can use a scheme called \emph{regularization by dimension reduction} (RDR)~\cite{Siegel:1979wq,CAPPER1980479,Avdeev:1982xy,HOWE1984409}. In this scheme, the $\mathcal{D}$-dimensional space is decomposed into a direct sum of $D$- and ($\mathcal{D}-D$)-dimensional subspaces. One can obtain $\mathcal{N}=1$ in $D=8$, $\mathcal{N}=2$ in $D=6$, $\mathcal{N}=4$ in $D=4$ and $\mathcal{N}=8$ in $D=2$ theories starting from $\mathcal{N}=1$ with $\mathcal{D}_{\textrm{max}}=10$. One can obtain $\mathcal{N}=2$ in $D=4$ and $\mathcal{N}=4$ in $D=2$ starting from $\mathcal{N}=1$ with $\mathcal{D}_{\textrm{max}}=6$, while one can obtain $\mathcal{N}=2$ in $D=2$ starting from $\mathcal{D}_{\textrm{max}}=4$~\cite{Brink:1976bc,VazquezMozo:1999ic}. In the following, we will use $d$ to represent the dimension of the momentum in all theories.

The evaluation of Feynman diagrams for theories that are obtained by dimensional reduction of $\symod$ can be carried out in two equivalent ways, one is working throughout in $(D-2\epsilon)\oplus (\mathcal{D}-D+2\epsilon)$ space for the theories we target, with $\epsilon$ being an infinitesimal quantity used for regularization and taking the dimension of the momentum-space to be $d=D-2\epsilon$. This prescription results in one having to introduce $\epsilon$-scalars into the Lagrangian in order to preserve supersymmetry \cite{CAPPER1980479}.  A simpler way to maintain supersymmetry is to take all fields in (\ref{actN1}) to be $\mathcal{D}$-dimensional tensors or spinors and all momentum to be $d=D-2\epsilon$ vectors.  We will use this second RDR scheme since it is the most transparent and efficient for computations \cite{Siegel:1979wq}.

With this in mind, up to thermal mass corrections, for any SYM theory that is characterized by $(\mathcal{D}_{\textrm{max}}, d)$, massless (unresummed) contributions to the free energy can be calculated perturbatively by computing vacuum Feynman diagrams using $\mathcal{N}=1$ SYM in $\mathcal{D}=\mathcal{D}_{\textrm{max}}$ and then restricting the momentum in loops to $d$ dimensions to obtain the result in the target theory.

\section{$\mathcal{N}=4$ supersymmetric Yang-Mills theory in $4$-dimensions ($\symff$)}
\la{N4_theory}

The $\symff$ theory can be obtained by dimensional reduction of $\symod$ in \mbox{$\mathcal{D}=\mathcal{D}_{\textrm{max}}=10$} with all fields being in the adjoint representation of $SU(N_c)$. By imposing the Majorana-Weyl condition, the spinor in 10 dimensions can be expressed as four Majorana spinors
\be
\psi_i \equiv  \begin{pmatrix} \lambda_{i,\alpha}\\  \bar{\lambda}^{\dot{\alpha}}_i \end{pmatrix}  \quad\quad \textrm{and} \quad\quad \bar{\psi}_i \equiv  \begin{pmatrix} \lambda^\alpha_i & \bar{\lambda}_{i,\dot{\alpha}} \end{pmatrix} ,
\ee
where $i=1,2,3,4$ indexes the Majorana fermions and $\psi_i$ denotes each bispinor. This allows one to convert the two-component Weyl spinors $\lambda$ in four dimensions into four-component Majorana fermions which satisfy $\bar{\lambda}^{\dot{\alpha}}\equiv [\lambda^\alpha]^\dagger$, $\alpha=1,2$ \cite{Quevedo:2010ui,bertolini2015lectures,Yamada:2006rx,DHoker:1999yni,Kovacs:1999fx}. The conjugate spinor $\bar{\psi}$ is not independent, but is related to $\psi$ via the Majorana condition $\psi=C\bar{\psi}$, where $C=\begin{pmatrix}\begin{smallmatrix} \epsilon_{\alpha\beta} & 0 \\ 0 & \epsilon^{\dot{\alpha}\dot{\beta}}\end{smallmatrix} \end{pmatrix}$ is the charge conjugation operator with $\epsilon_{02}=-\epsilon_{11}\equiv-1$. The fermionic fields can be expanded in the basis of color generators as $\psi_i=\psi_i^a t^a$. The coefficients $\psi_i^a$ are four-component Grassmann-valued spinors.

By decomposing the vector field in 10 dimensions, one obtains six independent real scalar fields which are represented by a multiplet
\ba\la{la}
 \Phi\equiv (X_1,Y_1,X_2,Y_2,X_3,Y_3) \, ,
\ea
where $X_{\texttt{p}}$ and $Y_{\texttt{q}}$ are Hermitian, with ${\texttt{p,q}}=1,2,3$. $X_{\texttt{p}}$ and $Y_{\texttt{q}}$ denote scalars and pseudoscalar fields, respectively. We will use a capital Latin index $A$ to denote components of the vector $\Phi$. Therefore $\Phi_A$, $X_{\texttt{p}}$, and $Y_{\texttt{q}}$ can be expanded as $\Phi_A=\Phi_A^a t^a$, with $A=1, \cdots ,6$, and $X_{\texttt{p}}=X_{\texttt{p}}^at^a$, $Y_{\texttt{q}}=Y_{\texttt{q}}^at^a$.

The action and Lagrangian that generates the perturbative expansion for $\symff$ in Minkowski-space can be expressed as
\ba\la{lag}
S_{\symff}&=& \int d^4 x \, \mathcal{L}_{\symff} \; , \quad\quad \textrm{with}\nonumber \\
\mathcal{L}_{\symff}&=& \textrm{Tr}\bigg[{-}\frac{1}{2}G_{\mu\nu}^2+(D_\mu\Phi_A)^2+i\bar{\psi}_i {\displaystyle{\not} D}\psi_i-\frac{1}{2}g^2(i[\Phi_A,\Phi_B])^2 \nonumber \\
&& \hspace{1cm} - i g \bar{\psi}_i\big[\alpha_{ij}^{\texttt{p}} X_{\texttt{p}}+i \beta_{ij}^{\texttt{q}}\gamma_5Y_{\texttt{q}},\psi_j\big] \bigg] +\mathcal{L}_{\textrm{gf}}+\mathcal{L}_{\textrm{gh}}+\Delta\mathcal{L}_{\textrm{SYM}} \, ,
\ea
where $\mu, \nu=0,1,2,3$ and $\alpha^{\texttt{p}}$ and $\beta^{\texttt{q}}$ are $4\times 4$ matrices that satisfy
\ba\la{lag1}
\{\alpha^{\texttt{p}},\alpha^{\texttt{q}}\}=-2\delta^{\texttt{pq}} \, , \quad \{\beta^{\texttt{p}},\beta^{\texttt{q}} \}=-2\delta^{\texttt{pq}} \, , \quad [\alpha^{\texttt{p}},\beta^{\texttt{q}}]=0 \, .
\ea
Up to unitary transformation, their explicit form is
\ba\la{lag1x}
&&\alpha^1=\begin{pmatrix} 0& \sigma_1 \\ -\sigma_1 &0 \end{pmatrix},  \quad \quad \alpha^2=\begin{pmatrix} 0& -\sigma_3 \\ \sigma_3 &0 \end{pmatrix},  \quad \quad  \alpha^3=\begin{pmatrix} i\sigma_2& 0 \\ 0&i\sigma_2 \end{pmatrix},  \nonumber \\&& \beta^1=\begin{pmatrix} 0& i\sigma_2 \\ i\sigma_2 &0 \end{pmatrix},  \quad \quad  \beta^2=\begin{pmatrix} 0& \sigma_0 \\ -\sigma_0 &0 \end{pmatrix},  \quad \quad  \beta^3=\begin{pmatrix} -i\sigma_2&0 \\ 0 & i\sigma_2 \end{pmatrix} ,
\ea
where $\sigma_i$ with $i \in \{1,2,3\}$ are the $2\times 2$ Pauli matrices. The matrices $\alpha$ and $\beta$ satisfy $\alpha_{ik}^{\texttt{p}}\alpha_{kj}^{\texttt{p}}=-3\delta_{ij}$ and $\beta_{ij}^{\texttt{q}}\beta_{ji}^{\texttt{p}}=-4\delta^{\texttt{pq}}$, with $\delta_{ii}=4$ for four Majorana fermions and $\delta^{\texttt{pp}}=3 $ for three scalars.

The ghost term $\mathcal{L}_{\textrm{gh}}$ depends on the choice of the gauge-fixing term $\mathcal{L}_{\textrm{gf}}$ and is the same as in QCD.  Here we work in general covariant gauge, giving
\ba\la{gf}
\mathcal{L}_{\textrm{gf}}^{\symff} &=& -\frac{1}{\xi}\textrm{Tr}\big[(\partial^\mu A_\mu)^2 \big],  \nonumber \\
\mathcal{L}_{\textrm{gh}}^{\symff} &=&-2\textrm{Tr}\big[\bar{\eta}\,\partial^\mu \! D_\mu \eta \big] ,
\ea
with $\xi$ being the gauge parameter.

Since we want to obtain the thermodynamic functions up to $\mathcal{O}(\lambda^{2})$, we need to calculate Feynman diagrams through three loop order. However at three loop level in QCD \cite{Arnold:1994ps,Arnold:1994eb}, infrared divergences appear that need to be canceled by summing over the ring diagrams appearing in the thermal mass counterterm.  As a result, there will be non-analytic contributions to the free energy at order $\lambda^{3/2}$ and $\lambda^2 \log\lambda$, representing a breakdown of naive perturbation theory due to infrared thermal physics. In QCD, the need for such reorganizations of perturbation theory stems from the behavior of the thermal propagator at soft momentum, $p_\text{soft} \sim \sqrt{\lambda}T$.  In the weak-coupling limit, the temperature is much higher than $p_\text{soft}$ and one finds that the thermal mass cannot be treated as a perturbation in the static propagator of bosonic fields.

As detailed in refs.~\cite{Arnold:1994ps,Arnold:1994eb}, in order to systematically resum the necessary diagrams, we need to modify the static bosonic propagators by incorporating gluon and scalar thermal masses, $m_D$ and $M$, respectively.  Such a reorganization is necessary in order to make finite-temperature perturbation theory well-defined beyond $\mathcal{O}(\lambda)$. In practice, the thermal masses are determined at leading order by computing the gluon and scalar self-energies $\Pi_{\mu\nu}$ and ${\cal P}$ at zero four-momentum.  In this manner, a scalar thermal mass is generated for $\Phi$ and a gluonic thermal mass is generated for the $A_0$ but not for $\vec{\textbf{A}}$.  We will introduce $m_D$ and $M$ only for the zero Matsubara modes of the gluon and scalar fields, generalizing the method introduced by Arnold and Zhai~\cite{Arnold:1994ps,Arnold:1994eb}.  In the context of finite temperature QCD, such resummations have been carried out through ${\cal O}(\lambda^3 \log\lambda)$ \cite{Arnold:1994ps,Arnold:1994eb,Zhai:1995ac,Braaten:1995jr,Kajantie:2002wa} and reorganizations of perturbation theory based on hard thermal loop perturbation theory and the $\Phi$-derivable approach have shown that one can extend the radius of convergence of weak-coupling treatments to intermediate couplings $g \sim 2$ \cite{Andersen:1999fw,Blaizot:1999ap,Andersen:2002ey,Andersen:2003zk,Andersen:2010ct,Andersen:2011sf,Blaizot:2000fc}.  We note that such perturbative reorganizations have also been carried out to two-loop order in $\symff$ \cite{Blaizot:2006tk,Du:2020odw}.

Following Arnold and Zhai, in this work we introduce thermal masses, $m_D$ and $M$, only for the zero Matsubara modes of the gluon and scalar fields.  The resulting reorganized Lagrangian density in frequency space can be rewritten as
\ba\la{lagresum}
\mathcal{L}_{\symff}^{\textrm{resum}}&=& \{\mathcal{L}_{\symff}+ \textrm{Tr}[m_D^2 A_0^2\delta_{p_0} -M^2\Phi_A^2\delta_{p_0}]\}
-  \textrm{Tr}[m_D^2 A_0^2\delta_{p_0} -M^2\Phi_A^2\delta_{p_0}]\, ,
\ea
where $\delta_{p_0}$ is shorthand for the Kronecker delta function $\delta_{{p_0},0}$. Then we absorb the two $A_0^2$ and $\Phi^2 $ terms in the curly brackets into our unperturbed Lagrangian $\mathcal{L}_0$, and treat the two terms outside the curly brackets as a perturbation.

At the $n^\text{th}$ loop order in the Arnold and Zhai reorganization the leading term contributing to the thermodynamic potential is ${\cal O}(\lambda^{n-1})$; however, due to the thermal masses, each loop order can contain terms that are higher order in $\lambda$.  Such corrections make the map between loop order and the contributing perturbative orders more complicated.  Note importantly that it is not possible for the thermal mass corrections to generate lower-order terms, they only generate higher-order terms than ${\cal O}(\lambda^{n-1})$ at each loop order.  As a result, there can be no modification of the ${\cal O}(\lambda^2)$ contribution coming from four or higher-order loop graphs.  We also note that terms containing logarithms formally contribute at the same order as the power of $\lambda$ multiplying them, i.e. $\lambda^2$ and $\lambda^2 \log\lambda$ both contribute at ${\cal O}(\lambda^2)$.

Similar to QCD, in order to obtain the thermodynamics up to $\mathcal{O}(\lambda^{2})$, we need to calculate massive Feynman diagrams up to two-loop order and we can take the three-loop diagrams to have bare propagators.  Since the propagators in the three-loop diagrams are massless, one can use the RDR method to compute these contributions in $\symot$ and then using $g^M_M=10$, $\mathcal{D}_{\textrm{max}}=10$, and $\textrm{Tr}\,I_n=8$, with all momentum integrals computed in $d=4-2\epsilon$.  This dramatically reduces the number of diagrams that one must evaluate.  For the massive two-loop diagrams, however, we must compute directly in $\symff$, since we must dress the gluon and scalar propagators differently.

As explained in ref.~\cite{Arnold:1994ps,Arnold:1994eb}, the one- and two-loop free energies are gauge parameter independent in the resummed scheme in QCD. In $\symff$, one has additional scalars and fermions, however, their propagators do not depend on the gauge parameter. Since the $\symff$ gluon propagator has the same form as in QCD (up to the definition of the gluon thermal mass $m_D$), the one-loop free energies are still gauge invariant.  In addition, we will demonstrate that the two-loop diagrams which contain a three/four-gluon vertex, a gluon-ghost vertex, and/or gluon-quark vertex are gauge invariant in $\symff$.

\section{NNLO thermodynamics of $\symff$}\la{potential}

Since the RDR method is based on analytically continuing only the number of coordinates and the momentum, the regularized momentum integral can be defined in a similar manner as in QCD. Since we are considering $\symff$, the regularized momentum integral in $d=4-2\epsilon$ dimensions can be defined as
\ba\la{integraln4}
\sumint_P & \equiv & \mu^{2\epsilon} T \sum_{P_0=2n \pi T}\int\frac{d^{d-1}p}{(2\pi)^{d-1}} \, ,\\
\sumint_{\{P\}}  & \equiv &   \mu^{2\epsilon} T \sum_{P_0=(2 n+1) \pi T }\int\frac{d^{d-1}p}{(2\pi)^{d-1}} \,  ,
\ea
where $\mu^{2\epsilon} = \big(\frac{e^{\gamma_E}\overline{\mu}^2}{4\pi}\big)^\epsilon$ with $\bar{\mu}$ being an arbitrary momentum scale.

It is also convenient to introduce various invariants associated with the representations of the SU($N_c$) gauge group.  Denoting the generators of the adjoint representation as $(T^a)^{bc}=-i f^{abc}$ and generators of the fundamental representation as $t^a$ we define the following group theory factors
\ba\la{generator}
{\rm Tr}\,T^a T^b &=& c_A \, \delta^{ab} \, , \nonumber \\
\delta^{aa} &=& d_A \, , \nonumber \\
\left[t^a t^a\right]_{ij} &=& c_F \, \delta_{ij} \, , \nonumber \\
{\rm Tr}\,t^a t^b &=& \half \, \delta^{ab} \, .
\ea
With the standard normalization
\ba
c_A &=& N_c \nonumber \, , \\
d_A &=& N_c^2-1 \nonumber \, .
\ea

\subsection{The resummed one-loop free energy}
\label{sect:oneloop}

\begin{figure}[t]
\begin{center}
\includegraphics[width=0.55\linewidth]{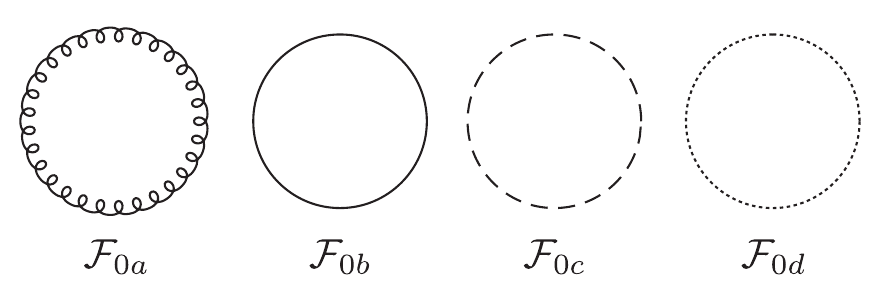}
\end{center}
\vspace{-5mm}
\caption{One-loop diagrams contributing to the $\symff$ free energy. Spiral lines indicate gluons, solid lines indicate Majorana fermions (gauginos), dashed lines indicate scalars, and dotted lines indicate ghosts. }
\label{fig:diagrams1}
\end{figure}

For the one-loop contributions, using the resummed Lagrangian density \eqref{lagresum}, one finds that the one-loop free energy for gluons and scalars will result in a perturbative contribution at ${\cal O}(\lambda^{3/2})$.  The one-loop free energy can be written as
\be
F_\text{1-loop}^{\textrm{resum}} = d_A  \mathcal{F}_{0a} +d_F\mathcal{F}_{0b}+d_S\mathcal{F}_{0c}+d_A \mathcal{F}_{0d} \, ,
\ee
where $d_A=N_c^2-1$ is the dimension of the adjoint representation. The Feynman diagrams corresponding to each term are presented in fig.~\ref{fig:diagrams1}. We note that there are four independent Majorana fermions in the adjoint representation giving $d_F=4d_A$.  For the scalars, one has $d_S=6d_A$.

Using the resummed gluonic propagator, $\mathcal{F}_{0a}$ can be expressed as
\ba\la{oneloopb}
\mathcal{F}_{0a} = \frac{1}{2}\sumint_P \big[ D\log P^2+\log(p^2+m_D^2) \big] = \frac{D}{2}b_0-\frac{m_D^3 T}{12\pi}\,,
\ea
where $D$ is the number gluon polarization states.  In vacuum, two of the polarizations are unphysical and are canceled by the ghost contribution. We note that this same form can be obtained from the one-loop gluonic hard-thermal-loop perturbation theory (HTLpt) result~\cite{Du:2020odw} by setting the gluonic transverse self-energy to zero and the longitudinal self-energy to $m_D^2$. The coefficient $b_0$ is defined as
\ba
b_0 \equiv \sumint_P \log P^2 = -\frac{\pi^2}{45} T^4.
\ea
Making use of the resummed scalar propagator, ${\cal F}_{\rm 0c}$ can be expressed as
\be\la{oneloops}
{\cal F}_{\rm 0c} = \frac{1}{2}\sumint_P \{\log P^2+\log(p^2+M^2) \} = \frac{1}{2}b_0-\frac{M^3 T}{12\pi} \, .
\ee
Since there are no thermal mass contributions for fermions and ghosts, their one-loop free energies have simpler forms
\ba
{\cal F}_{\rm 0b} &=& -f_0 \, , \la{oneloopf}\\
{\cal F}_{0d}  &=& -b_0 \, , \la{oneloopg}
\ea
with $f_0$ defined as
\ba
f_0 \equiv \sumint_{\{P\}} \log P^2 = \frac{7\pi^2}{360} T^4.
\ea
We note that the resulting one-loop gluon and scalar free energies are equal to the corresponding summation of the leading-order one-loop hard and soft contributions in HTLpt when taking $m_q \rightarrow 0$~\cite{Du:2020odw}. By combining (\ref{oneloopb}), (\ref{oneloops}), (\ref{oneloopf}), and (\ref{oneloopg}) one obtains
\be
F_\text{1-loop}^{\textrm{resum}} = d_A \bigg[\frac{D+4}{2}b_0-4f_0-\frac{T}{12\pi}(m_D^3+6M^3) \bigg]  \, .
\ee
By imposing $D=4$, $m_D^2=2\lambda T^2$, $M^2=\lambda T^2$, and truncating at $\mathcal{O}(\epsilon^0)$ one obtains
\be\la{1loopresum}
F_\text{1-loop}^{\textrm{resum}} =-d_A \bigg(\frac{\pi^2 T^4 }{6}\bigg) \bigg[ 1 +\frac{3+\sqrt{2}}{\pi^3} \lambda^{3/2} \bigg]  \, .
\ee
%

\subsection{The resummed two-loop free energy}
\label{sect:twoloop}

\begin{figure}[t]
\begin{center}
\includegraphics[width=0.85\linewidth]{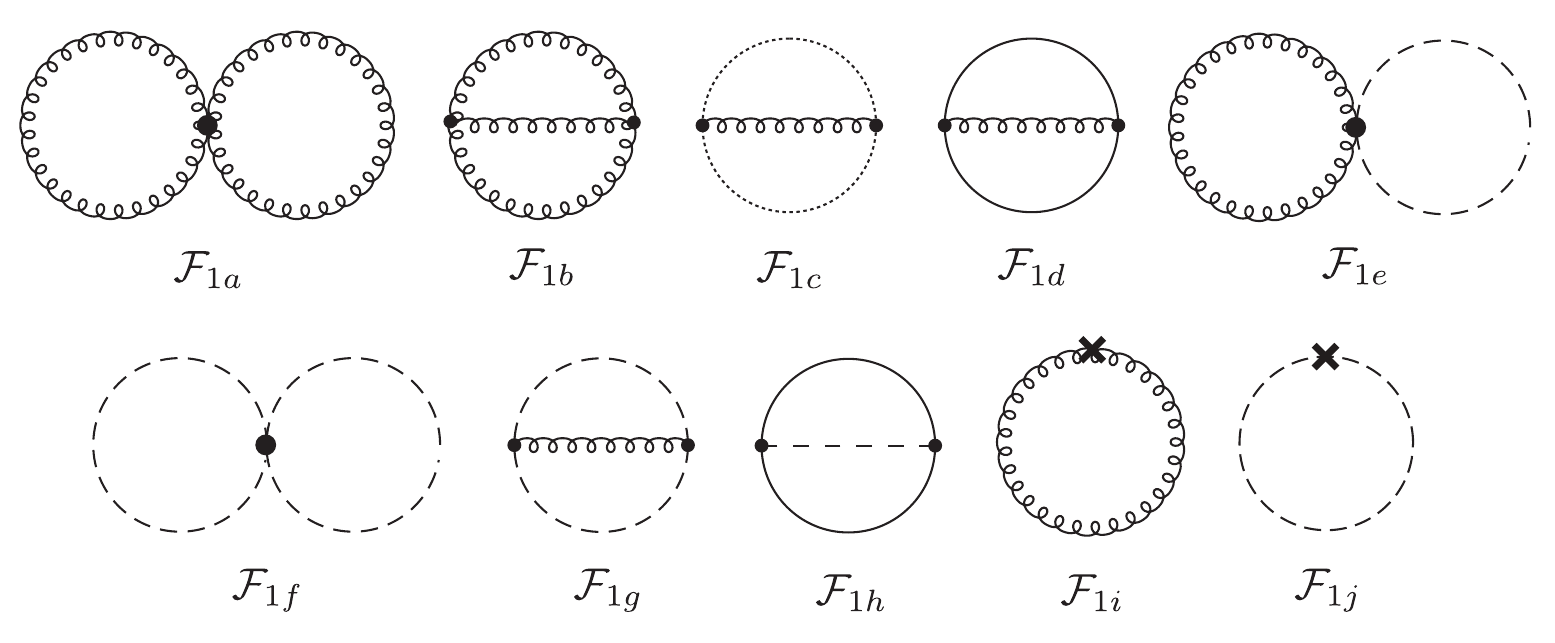}
\end{center}
\vspace{-4mm}
\caption{Two-loop contributions to the  $\symff$ free energy. The crosses are the thermal counterterms produced by the last two terms of (\ref{lagresum}).}
\label{fig:diagrams2}
\end{figure}

The $\symff$ two-loop free energy can be written as
\be
F_\text{2-loop}^{\textrm{resum}} = d_A \bigg\{\lambda [{\cal F}_{1a}+{\cal F}_{1b}+{\cal F}_{1c}+{\cal F}_{1d}+{\cal F}_{1e}+{\cal F}_{1f}+{\cal F}_{1g}+{\cal F}_{1h}]+{\cal F}_{1i}+{\cal F}_{1j}  \bigg\} \, .
\ee
The Feynman diagrams corresponding to each term are presented in fig.~\ref{fig:diagrams2}.  Making use of the resummed gluon and scalar propagators, one obtains
\ba
\mathcal{F}_{1a}&=&   \frac{D(D-1)}{4} b_1^2 - \frac{m_DT}{8\pi}  (D-1)b_1      \,, \la{2loopn41}\\
\mathcal{F}_{1b} &=&   -\frac{3}{4}(D-1) b_1^2 +\delta_1+\delta_2     \,,\la{2loopn42}\\
\delta_1 &=&  m_D^2\sumint_{PQ}\frac{\delta_{p_0}(1-\delta_{q_0})}{(P^2+m_D^2)Q^2(P+Q)^2}-(D-\frac{3}{2})\textrm{I}_{\textrm{resum}}^b  - \frac{m_DT}{8\pi } \bigg(D-\frac{5}{2} \bigg)b_1  \, ,\nonumber \\
\delta_2 &=& \frac{1}{4}\frac{m_D^2T^2}{(4\pi)^2}+ m_D^2\sumint_{PQ}\frac{\delta_{p_0}\delta_{q_0}}{(P^2+m_D^2)(Q^2+m_D^2)(P+Q)^2} \, , \nonumber \\
\mathcal{F}_{1c}&=&    \frac{1}{4} b_1^2 +\frac{1}{2} \textrm{I}_{\textrm{resum}}^b  - \frac{m_DT}{8\pi} \big(-\frac{1}{2}b_1  \big)       \,, \la{2loopn43}\\
\mathcal{F}_{1d} &=&     -2(D-2)f_1(2 b_1-f_1)+8 \textrm{I}_{\textrm{resum}}^f-2m_D^2\sumint_{P} \frac{\delta_{p_0}}{P^2}\Pi^f(P) - \frac{m_DT}{8\pi} 8 b_1    \,,\la{2loopn44}\\
\mathcal{F}_{1e}&=&    3D b_1^2 +3\frac{M m_D T^2}{(4\pi)^2} - \frac{M T}{8\pi} 6 D b_1 -  \frac{m_D T}{8\pi} 6 b_1       \,,\la{2loopn45}\\
\mathcal{F}_{1f}&=&   \frac{15}{2}b_1^2 +\frac{15}{2}\frac{M^2 T^2}{(4\pi)^2} - \frac{M T}{8\pi} 30 b_1   \,,\la{2loopn46}\\
\mathcal{F}_{1g} &=&     -\frac{9}{2}b_1^2 +\delta_3 +\delta_4  \,,\la{2loopn47}\\
\delta_3 &=& 6 M^2\sumint_{PQ}\frac{\delta_{p_0}(1-\delta_{q_0})}{(P^2+M^2)Q^2(P+Q)^2} -6 \textrm{I}_{\textrm{resum}}^b - \frac{m_D T}{8\pi  }6 b_1 - \frac{M T}{8\pi  }(-6) b_1\, ,\nonumber  \\
\delta_4 &=& \frac{3}{2}\frac{M^2 T^2}{(4\pi)^2} + 6 M^2\sumint_{PQ}\frac{\delta_{p_0}\delta_{q_0}}{(P^2+M^2)(Q^2+M^2)(P+Q)^2} \,,\nonumber \\
\mathcal{F}_{1h}&=&    -12f_1(2 b_1-f_1) - 12 M^2 \sumint_{P} \frac{\delta_{p_0}}{P^2}\Pi^f(P) - \frac{M T}{8\pi} 24b_1  \, , \la{2loopn48}\\
\mathcal{F}_{1i} &=& \frac{m_D T}{8\pi}m_D^2 \, ,\la{2loopn49}\\
\mathcal{F}_{1j}&=& \frac{M T}{8\pi}6 M^2 \la{2loopn410}\,,
\ea
where
\ba\la{forluma1x}
b_1 &=& \frac{T^2}{12}\bigg[ 1+ \epsilon \bigg( 2\log\frac{\bar{\mu}}{4\pi T} +2 \frac{\zeta^{'}(-1)}{\zeta(-1)} +2 \bigg) \bigg] +\mathcal{O}(\epsilon^2) \, , \nonumber\\
b_2 &=& \frac{1}{(4\pi)^2}\bigg[ \frac{1}{\epsilon} +2 \log\frac{\bar{\mu}}{4\pi T} +2\gamma_E \bigg] +\mathcal{O}(\epsilon) \, .
\ea
The fundamental integrals appearing above are defined as
\ba\la{forluma1}
b_n \equiv \sumint_{P} \frac{1}{P^{2n}} \, , \quad\quad f_n \equiv \sumint_{\{P\}} \frac{1}{P^{2n}} =(2^{2n+1-d}-1)b_n \, ,  \quad n \geq 1\,,
\ea
and
\ba
\Pi^f(P) &\equiv& \sumint_{\{Q\}}\frac{1}{Q^2(P+Q)^2}\,, \la{forluma1xx1} \\
\textrm{I}_{\textrm{resum}}^b &\equiv & \sumint_{PQ}\bigg[ \frac{\delta_{p_0}}{P^2+m_D^2}-\frac{\delta_{p_0}}{P^2} \bigg] \bigg[ \frac{Q_0^2}{Q^2(P+Q)^2} -\frac{Q_0^2}{Q^4}\bigg]  \nonumber\\ &=& - m_D^2\sumint_{PQ}\frac{\delta_{P_0}}{P^4} \bigg[ \frac{Q_0^2}{Q^2(P+Q)^2}-\frac{Q_0^2}{Q^4}  \bigg]+\mathcal{O}(g^3) \, ,\la{forluma1xx2}\\
\textrm{I}_{\textrm{resum}}^f &\equiv& \sumint_{P\{Q\}}\bigg[ \frac{\delta_{p_0}}{P^2+m_D^2}-\frac{\delta_{p_0}}{P^2} \bigg] \bigg[ \frac{Q_0^2}{Q^2(P+Q)^2} -\frac{Q_0^2}{Q^4}\bigg]\la{forluma1xx3}\,.
\ea

One finds that the contributions $\mathcal{F}_{1a}, \mathcal{F}_{1b}$, and $\mathcal{F}_{1c}$ are the same as in QCD \cite{Arnold:1994ps} and that $\mathcal{F}_{1d}$ is two times the result obtained in QCD in \cite{Arnold:1994eb}, up to the definition of $m_D$ and $D$. As in QCD, the two-loop free energy is gauge independent using our resummation method, so that $\mathcal{F}_{1d}$ and the summation of $\mathcal{F}_{1a}, \mathcal{F}_{1b}$ and $\mathcal{F}_{1c}$ are gauge parameter independent.  The only remaining contributions which depend on the gauge parameter are $\mathcal{F}_{1e}$ and $\mathcal{F}_{1g}$.  By using the full propagator in \ref{gpropagatpr}, the gauge terms in $\mathcal{F}_{1e}$ and $\mathcal{F}_{1g}$ are $-3(1-\xi)(b_1^2-\frac{MT}{4\pi}b_1)$ and  $3(1-\xi)(b_1^2-\frac{MT}{4\pi}b_1)$, respectively, and hence cancel.  This proves that our resummed two-loop result is gauge-independent in the RDR scheme.

The form of $\mathcal{F}_{1g}$ is similar to $\mathcal{F}_{1b}$. The first term in each is the result when $m_D$ and $M$ are set to zero.  The terms containing $\delta_1$ and $\delta_3$ are the corrections in which only one of the three momenta possesses a zero mode (hard-hard-soft).  The terms containing $\delta_2$ and $\delta_4$ stem from contributions in which all the three momenta possess zero modes (soft-soft-soft). For $\mathcal{F}_{1g}$
\ba
\delta_3 &=&\frac{3}{2}\sumint_{PQ} \bigg\{ \frac{\delta_{p_0}(1-\delta_{q_0})}{P^2+M^2}  \frac{-2(2P+Q)^2}{Q^2(P+Q)^2} +\bigg[ \frac{\delta_{p_0}}{P^2+m_D^2}-\frac{\delta_{p_0}}{P^2} \bigg] \frac{-4Q_0^2}{Q^2(P+Q)^2}  \bigg\} \nonumber \\ & & \hspace{1cm} - \{m_D \rightarrow 0,M\rightarrow 0\}  \, ,
\ea
%
where the quantity in brackets on the second line corresponds to the sum-integral on the first line, but with thermal masses taken to zero. The factors of $2P+Q$ and $Q_0$ appearing in the expression come from the gluon-scalar vertex. Making use of the relation $(2P+Q)^2=2P^2+2(P+Q)^2-Q^2$, $\delta_3$ can be reduced to
\ba
\delta_3 &=&6 M^2\sumint_{PQ}\frac{\delta_{p_0}(1-\delta_{q_0})}{(P^2+M^2)Q^2(P+Q)^2} -6\sumint_{PQ}\bigg[ \frac{\delta_{p_0}}{P^2+m_D^2}-\frac{\delta_{p_0}}{P^2} \bigg] \frac{Q_0^2}{Q^2(P+Q)^2} \nonumber \\ & & -3 b_1 \sumint_{P} \frac{\delta_{p_0}}{P^2+M^2}     \, ,
\ea
which can be written in the same form as in eq.~\eqref{2loopn47}.   An expression for $\delta_4$ can be calculated similarly.

The form of $\mathcal{F}_{1d}$ and $\mathcal{F}_{1h}$ are similar with the exception of the appearance of $\textrm{I}_{\textrm{resum}}^f$.  This is due to the tensor nature of the gluon propagator.  The contribution $\mathcal{F}_{1h}$ can be written as
\ba
\mathcal{F}_{1h} &=&  -24  \sumint_{P\{Q\}}\frac{Q^2+P\cdot Q}{Q^2(P+Q)^2} \bigg[ \frac{1}{P^2}+\bigg(  \frac{1}{P^2+M^2}-\frac{1}{P^2} \bigg)\delta_{p_0}   \bigg]  ,
\ea
where $Q^2+P\cdot Q=\frac{1}{2}[(P+Q)^2+Q^2-P^2]$. It can be reduced to
\ba\la{f1hx}
\mathcal{F}_{1h} &=&  -12 \sumint_{P\{Q\}} \frac{2}{P^2Q^2}+12\sumint_{\{P Q\}}\frac{1}{P^2Q^2}-24 \sumint_{P\{Q\}}\frac{\delta_{p_0}}{(P^2+M^2)Q^2} \nonumber \\ &&\hspace{1cm}-12M^2\sumint_{P\{Q\}}\frac{\delta_{p_0}}{(P^2+M^2)Q^2(P+Q)^2}\,.
\ea
Since the last term in eq.~(\ref{f1hx}) is divergence free, one can ignore the $M^2$ in the denominator. As a result, the above equation can be written in the same form as eq.~\eqref{2loopn48}.

From ref.~\cite{Arnold:1994ps,Arnold:1994eb} one has
\ba\la{forluma2}
\textrm{I}_{\textrm{resum}}^b = -\frac{1}{8}\frac{m_D^2T^2}{(4\pi)^2} + \mathcal{O}(g^3,\epsilon) \,, \quad    \textrm{I}_{\textrm{resum}}^f = \mathcal{O}(g^3,\epsilon)\, ,
\ea
and
\ba
&& \sumint_{P} \frac{\delta_{p_0}}{P^2}\Pi^f(P) = -\frac{T^2}{(4\pi)^2}\log2 +\mathcal{O}(\epsilon)\,, \la{forluma31}\\
&& \sumint_{PQ}\frac{\delta_{p_0}(1-\delta_{q_0})}{(P^2+m^2)Q^2(P+Q)^2} = \frac{T^2}{(4\pi)^2} \bigg[  -\frac{1}{4\epsilon} +\log\frac{2T}{\overline{\mu}} -\frac{1}{2} \bigg]+\mathcal{O}(m,\epsilon) \, ,\la{forluma32}\\
&& \sumint_{PQ}\frac{\delta_{p_0}\delta_{q_0}}{(P^2+m^2)(Q^2+m^2)(P+Q)^2} = \frac{T^2}{(4\pi)^2} \bigg[  \frac{1}{4\epsilon} +\log\frac{\overline{\mu}}{2m} +\frac{1}{2} \bigg] +\mathcal{O}(\epsilon) \la{forluma33}\, .
\ea


One finds that the two-loop contributions proportional to $\lambda^{3/2}$ cancel since $m_D^2=24b_1 \lambda$ and $M^2=12b_1\lambda$. Combining all contributions, we obtain the two-loop resummed free energy
\ba\la{f4s1}
F_\text{2-loop}^{\textrm{resum}} & = & \lambda d_A  \bigg\{ \frac{D+4}{4}\bigg[(D+4)b_1^2-16b_1 f_1+8f_1^2 \bigg] +  6\frac{M^2T^2}{(4\pi)^2}\bigg( \frac{3}{2}-\log\frac{M}{T}+2 \log 2\bigg)  \nonumber \\ &&
\hspace{10mm} +\frac{m_D^2T^2}{(4\pi)^2}  \bigg( \frac{3}{4}+\frac{D}{8}-\log\frac{m_D}{T} +2 \log 2 \bigg)      +3\frac{m_D M T^2}{(4\pi)^2} \bigg\} \, .
\ea
Setting $D=4$, $m_D^2=2\lambda T^2, M^2=\lambda T^2$, and truncating at $\mathcal{O}(\epsilon^0)$, we obtain our final expression for the resummed two-loop contribution to the $\symff$  free energy
\ba\la{2loopresum}
F_\text{2-loop}^{\textrm{resum}} & = & -d_A \bigg(\frac{\pi^2 T^4}{6}\bigg) \bigg[ -\frac{3}{2\pi^2} \lambda-\frac{3}{2\pi^4} \bigg(  \frac{23}{8}+\frac{3\sqrt{2}}{4} +\frac{15 \log 2}{4} -  \log \lambda\bigg)\lambda^2 \bigg]\,.
\ea
%

\begin{figure}[t]
\begin{center}
\includegraphics[width=0.8\linewidth]{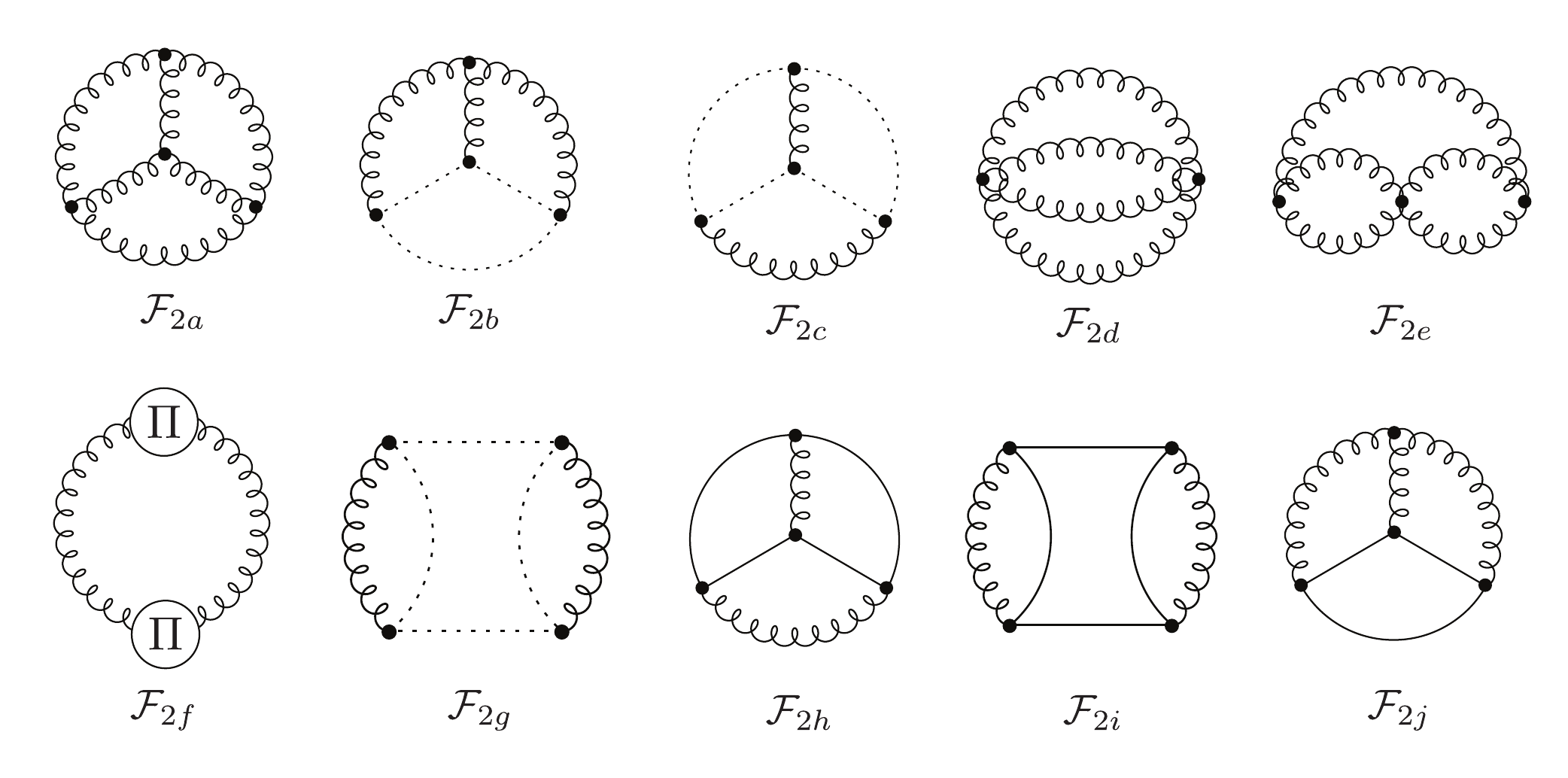}
\end{center}
\vspace{-7mm}
\caption{Three-loop vacuum diagrams contributing to the $\symot$ free energy.}
\label{fig:diagrams3}
\end{figure}

\begin{figure}[t]
\vspace{4mm}
\begin{center}
\includegraphics[width=0.5\linewidth]{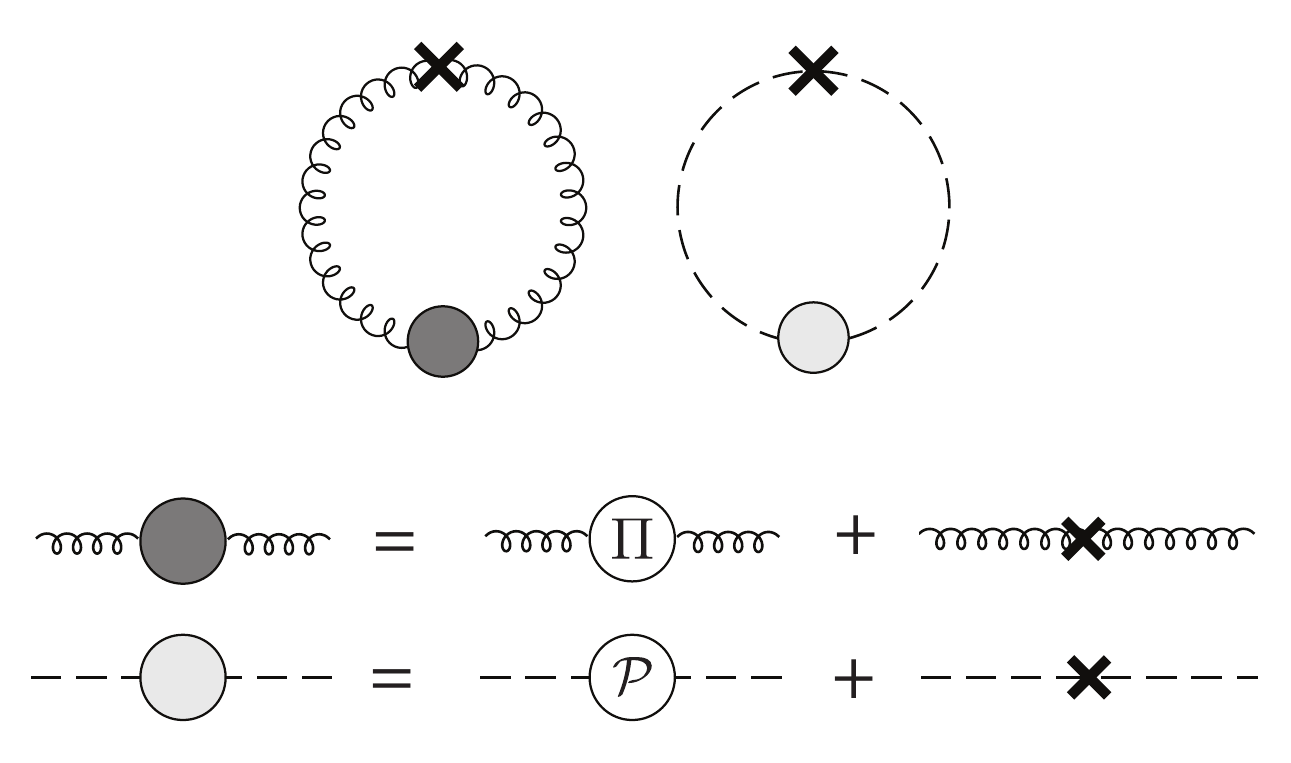}
\end{center}
\vspace{-4mm}
\caption{Three-loop gluon and scalar counterterm diagrams in $\symff$. The crosses are the ``thermal counter-terms'' arising from the last two terms of (\ref{lagresum}). Expressions for the $\symff$ gluon and scalar self-energies are presented in app.~\ref{self_energyn4}. }
\label{fig:diagrams4}
\end{figure}

\subsection{The resummed three-loop free energy}
\label{sect:treeloop}

As mentioned earlier, the calculation of the massless three-loop vacuum Feynman diagrams in $\symff$ can be accomplished more simply in the corresponding $\symot$ theory.  As a result of this equivalence, one can consider the much smaller set of $\symot$ graphs presented in fig.~\ref{fig:diagrams3}, which are topologically equivalent to three-loop QCD vacuum graphs.  The three-loop results in $\symff$  can be obtained by imposing $\mathcal{D} = \mathcal{D}_{\textrm{max}}=10$, $d=4-2\epsilon$ in the $\symod$ theory.
\be
F_\text{3-loop}^{\textrm{vacuum}} = d_A \lambda^2 [{\cal F}_{2a}+ {\cal F}_{2b}+{\cal F}_{2c}+{\cal F}_{2d}+{\cal F}_{2e}+{\cal F}_{2f}+{\cal F}_{2g} +{\cal F}_{2h}+{\cal F}_{2i} +{\cal F}_{2j}]|_{d=4-2\epsilon}^{{\cal D}=10}\,.
\ee
Using the Feynman rules in app.~\ref{propagatorn1}, one obtains
\ba
\mathcal{F}_{2a}&=& \bigg[-\frac{5\mathcal{D}}{8} +\frac{23}{32} \bigg] I_{\textrm{ball}}^{\textrm{bb}} \, ,  \la{f3loopn11}  \\
\mathcal{F}_{2b}&=& \frac{1}{16} I_{\textrm{ball}}^{\textrm{bb}} \, , \la{f3loopn12}\\
\mathcal{F}_{2c}&=&\frac{1}{32} I_{\textrm{ball}}^{\textrm{bb}}\, ,  \la{f3loopn13}\\
\mathcal{F}_{2d}&=&-\frac{3}{16}\mathcal{D}(\mathcal{D}-1) I_{\textrm{ball}}^{\textrm{bb}}\, , \la{f3loopn14} \\
\mathcal{F}_{2e}&=&\frac{27}{16}(\mathcal{D}-1) I_{\textrm{ball}}^{\textrm{bb}}\, , \la{f3loopn15}  \\
\mathcal{F}_{2g}&=& \frac{1}{8} I_{\textrm{ball}}^{\textrm{bb}} \, ,  \la{f3loopn16} \\
\mathcal{F}_{2f}&=& -\frac{1}{4} \big[ I_{\symod}^{\textrm{bb}} +I_{\symod}^{\textrm{bf}} +I_{\symod}^{\textrm{ff}} \big]
\, ,  \la{f3loopn17}\\
\mathcal{F}_{2h}&=& \frac{(\mathcal{D}-2)}{8}\textrm{Tr}\,I_n \bigg[ \frac{\mathcal{D}-6}{2}I_{\textrm{ball}}^{\textrm{ff}} + (4-\mathcal{D})I_{\textrm{ball}}^{\textrm{bf}}  \bigg] ,\la{f3loopn18}\\
\mathcal{F}_{2j}&=& -\frac{\mathcal{D}-2}{4}\textrm{Tr}\,I_n I_{\textrm{ball}}^{\textrm{bf}}\, , \la{f3loopn19}\\
\mathcal{F}_{2i}&=& \frac{(\mathcal{D}-2)^2}{4}\textrm{Tr}\,I_n \big[ I_{\textrm{ball}}^{\textrm{bf}}-2 H_3 + f_2 (f_1-b_1)^2 \big] ,  \la{f3loopn110}\,
\ea
where in all contributions we have taken $d=4-2\epsilon$.  Above
\ba
I_{\textrm{ball}}^{\textrm{bb}} &=& \sumint_{P} \big[\Pi^b(P)\big]^2, \la{Iballbb}\\
I_{\textrm{ball}}^{\textrm{bf}} &=& \sumint_{P} \Pi^b(P)\Pi^f(P) \, , \la{Iballbf}\\
I_{\textrm{ball}}^{\textrm{ff}} &=&\sumint_{P} \big[\Pi^f(P)\big]^2,  \la{Iballff}
\ea
with
\ba
\Pi^b(P) \equiv \sumint_{Q} \frac{1}{Q^2(P+Q)^2} \, ,
\ea
and
\ba
H_3&\equiv&  \sumint_{\{P\}QK} \frac{Q\cdot K}{P^2 Q^2 K^2(P+Q)^2 (P+K)^2} \, . \la{Hfun1}
\ea
The results for $I_{\textrm{ball}}^{\textrm{bb}}$, $I_{\textrm{ball}}^{\textrm{ff}}$, $I_{\textrm{ball}}^{\textrm{bf}}$, and $H_3$ truncated at ${\cal O}(\epsilon^0)$ have been given in refs.~\cite{Arnold:1994ps,Arnold:1994eb}
\ba
I_{\textrm{ball}}^{\textrm{bb}} &=& \frac{1}{(4\pi)^2}\bigg(\frac{T^2}{12} \bigg)^2 \bigg[\frac{6}{\epsilon} + 36\log\frac{\bar{\mu}}{4\pi T} -12 \frac{\zeta'(-3)}{\zeta(-3)}+48\frac{\zeta'(-1)}{\zeta(-1)} +\frac{182}{5} \bigg] +\mathcal{O}(\epsilon) \, , \la{Iballbbfinal} \\
I_{\textrm{ball}}^{\textrm{ff}} &=& \frac{1}{(4\pi)^2}\bigg(\frac{T^2}{12} \bigg)^2 \bigg[\frac{3}{2\epsilon} + 9\log\frac{\bar{\mu}}{4\pi T} -3 \frac{\zeta'(-3)}{\zeta(-3)}+12\frac{\zeta'(-1)}{\zeta(-1)} +\frac{173}{20}- \frac{63}{5} \log 2\bigg] +\mathcal{O}(\epsilon) \, , \la{Iballfffinal} \nonumber \\ \\
I_{\textrm{ball}}^{\textrm{bf}} &=& -\frac{1}{6} (1-2^{11-3d})I_{\textrm{ball}}^{\textrm{bb}}-\frac{1}{6}I_{\textrm{ball}}^{\textrm{ff}} \, ,\la{Iballbffinal}
\ea
and
\ba
H_3 = \frac{1}{(4\pi)^2}\bigg(\frac{T^2}{12} \bigg)^2 \bigg[\frac{3}{8\epsilon} &+ & \frac{9}{4}\log\frac{\bar{\mu}}{4\pi T} + \frac{3}{2} \frac{\zeta'(-3)}{\zeta(-3)}-\frac{3}{2}\frac{\zeta'(-1)}{\zeta(-1)} + \frac{9}{4} \gamma_E   \nonumber \\ &+ &\frac{361}{160} +\frac{57}{10} \log 2\bigg] +\mathcal{O}(\epsilon) \, . \la{H3}
\ea
The derivations of $I_{\symod}^{\textrm{bb}}$, $I_{\symod}^{\textrm{bf}}$, and $I_{\symod}^{\textrm{ff}}$ are presented in app.~\ref{In1bf}.

Infrared divergences will be generated by some of the diagrams in fig.~\ref{fig:diagrams3} due to three-momentum integrations involving massless gluon and scalar propagators in $d=4-2\epsilon$ dimensions. These divergences will be canceled by the thermal mass counterterm diagrams in fig.~\ref{fig:diagrams4}.   The first diagram in fig.~\ref{fig:diagrams4} represents the gluonic thermal counterterm contribution, where the shaded blob can be expressed as
\ba\la{shadedblobg}
\Delta \Pi_{\mu\nu}(P) \equiv \Pi_{\mu\nu}(P)-m_D^2\delta_{\mu 0}\delta_{\nu 0}\delta_{P_0} \, ,
\ea
and $\Pi_{\mu\nu}(P)=\Pi^b_{\mu\nu}(P)+\Pi^f_{\mu\nu}(P)$, as defined in eqs.~(\ref{self_energy_n4b1}) and (\ref{self_energy_n4b2}). The form of the second term is equivalent to the form presented in the ref.~\cite{Arnold:1994ps,Arnold:1994eb}, however, we choose the above form for clarity and calculational ease.  Similarly, the shaded blob in the second diagram in fig.~\ref{fig:diagrams4} stands for the scalar thermal counterterm and can be expressed as
\ba\la{shadedblobs}
\Delta\mathcal{P}(P) \equiv  \mathcal{P}(P)+ M^2 \delta_{P_0}  \, .
\ea
The first diagram in fig.~\ref{fig:diagrams4} is simple and is proportional to
\ba\la{diag4eq}
-\frac{1}{2}\sumint_P \frac{1}{P^4} m_D^2\Pi_{00}(P)\delta_{P_0} \, .
\ea
By inserting the one-loop self-energy presented in app.~\ref{self_energyn4} and using eq.~(\ref{forluma2}),
one can obtain the result in a straightforward manner. A similar procedure can be applied to the second diagram.

The final three-loop thermal mass counterterm contributions computed in $\symff$ are
\ba
\mathcal{F}_{\textrm{3-loop}}^{\textrm{sct}} &=&  d_A \lambda^2 6[(D+4)b_1-8f_1]\bigg[\sumint^{\,\prime}_{P} \frac{\Pi^b(P)}{P^2}-2\sumint^{\,\prime}_{P} \frac{\Pi^f(P)}{P^2}  \bigg]  \,, \la{f3loopn4counter1}\\
\mathcal{F}_{\textrm{3-loop}}^{\textrm{bct}} &=& d_A \lambda^2 (d-2) [(D+4)b_1-8f_1]
 \bigg[ \sumint^{\,\prime}_{P} \frac{\Pi^b(P)}{P^2}-2\sumint^{\,\prime}_{P} \frac{\Pi^f(P)}{P^2}   
-  \frac{1}{8} (D+4)  \frac{T^2}{(4\pi)^2}  \bigg] \la{f3loopn4counter2},
 \hspace{1cm}
\ea
where
\ba
\sumint^{\,\prime}_{P} \equiv \mu^{2\epsilon} T \sum_{P_0\neq 0} \int \frac{d^{3-2\epsilon}p}{(2\pi)^{3-2\epsilon}} \, .
\ea
To simplify further one can use eq.~(\ref{Isum})
\ba
\sumint^{\,\prime}_{P} \frac{\Pi^f(P)}{P^2} = - \sumint_{P} \frac{\delta_{p_0}}{P^2}\Pi^f(P)  \,,
\ea
and the following result from ref.~\cite{Arnold:1994ps}
\ba
\sumint^{\,\prime}_{P} \frac{\Pi^b(P)}{P^2} = \frac{T^2}{(4\pi)^2}\bigg[\frac{1}{4\epsilon} +\log\frac{\bar{\mu}}{4\pi T} +\log 2\pi + \frac{1}{2}\bigg]+\mathcal{O}(\epsilon)  \, .
\ea
Combining eqs.~(\ref{f3loopn11}) -- (\ref{f3loopn110}), (\ref{f3loopn4counter1}), and (\ref{f3loopn4counter2}), imposing $\mathcal{D}_{\textrm{max}}=10$, $D=4$, $\textrm{Tr}\,I_n=\mathcal{D}_{\textrm{max}}-2$, and inserting the results for all necessary integrals, through $\mathcal{O}(\epsilon^0)$ one obtains
\ba\la{3loopresum}
F_\text{3-loop}^{\textrm{resum}} &=& {\cal F}_\text{3-loop}^{\textrm{vacuum}}+\mathcal{F}_{\textrm{3-loop}}^{\textrm{sct}}+\mathcal{F}_{\textrm{3-loop}}^{\textrm{bct}} \nonumber\\& =& -d_A \bigg(\frac{\pi^2 T^4}{6} \bigg) \frac{\lambda^2}{2\pi^4} \bigg[  \frac{27}{8} +3\gamma+3 \frac{\zeta'(-1)}{\zeta(-1)} +5\log 2-6\log\pi \bigg]\,.
\ea
%

\section{$\symff$ thermodynamic functions to ${\cal O}(\lambda^2)$}\la{thermodynamics}

Combining eqs.~(\ref{1loopresum}), (\ref{2loopresum}), and (\ref{3loopresum}), we obtain our final result for the resummed free energy in the RDR scheme through ${\cal O}(\lambda^2)$
\ba
{\cal F} &=& -d_A \bigg(\frac{\pi^2 T^4}{6}\bigg) \bigg\{ 1-\frac{3}{2} \frac{\lambda}{\pi^2} +  \left( 3+\sqrt{2} \right) \left(\frac{\lambda}{\pi^2}\right)^{3/2}  \nonumber \\ && \hspace{1cm} + \bigg[ -\frac{21}{8} -\frac{9\sqrt{2}}{8} + \frac{3}{2} \gamma_E + \frac{3}{2}\frac{\zeta'(-1)}{\zeta(-1)}-\frac{25}{8} \log 2 + \frac{3}{2}\log \frac{\lambda}{\pi^2} \bigg] \left(\frac{\lambda}{\pi^2}\right)^2  \bigg \}. \hspace{8mm}
\label{eq:finalF}
\ea
We note that this result holds for all $N_c$.  As can be seen from this expression, one finds non-vanishing contributions at ${\cal O}(\lambda^2)$ and ${\cal O}(\lambda^2 \log\lambda)$, as anticipated.  Equation \eqref{eq:finalF} is manifestly finite due to an explicit cancellation between three-loop infrared singularities and the three-loop counterterm diagrams.  These cancellations remove all infrared divergent contributions.  In addition, there are no remaining poles due to ultraviolet divergences, since the coupling does not run in $\symff$ and, hence, no coupling constant renormalization counterterm is required.  Based on the result above, we see that the series organizes itself naturally as an expansion in $\lambda/\pi^2$, suggesting that the weak coupling expansion will break down for $\lambda \gtrsim \pi^2$.  The presence of the logarithm at order $\lambda^2$ stems directly from the dressing of the $A_0$ and scalar propagators.

The pressure, entropy density, and energy density can be obtained from \eqref{eq:finalF} using standard thermodynamic identities
\ba
\mathcal{P} &=& -\mathcal{F} \, , \nonumber \\
\mathcal{S} &=& -\frac{d \mathcal{F}}{d T}  \, , \nonumber \\
\mathcal{E} &=& \mathcal{F} - T\frac{d \mathcal{F}}{d T} \, .
\ea
Note that due the conformality of SYM, in all three of these functions, the only dependence on $T$ is contained in the overall factor of $\mathcal{F}_{\rm ideal}$.  As a result, when scaled by their ideal limits, the ratios of all of these quantities are the same, i.e. $\mathcal{P}/\mathcal{P}_{\rm ideal}$ = $\mathcal{S}/\mathcal{S}_{\rm ideal}$ = $\mathcal{E}/\mathcal{E}_{\rm ideal}$.

\begin{figure}[t]
\begin{center}
\vspace{-2mm}
\includegraphics[width=0.62\linewidth]{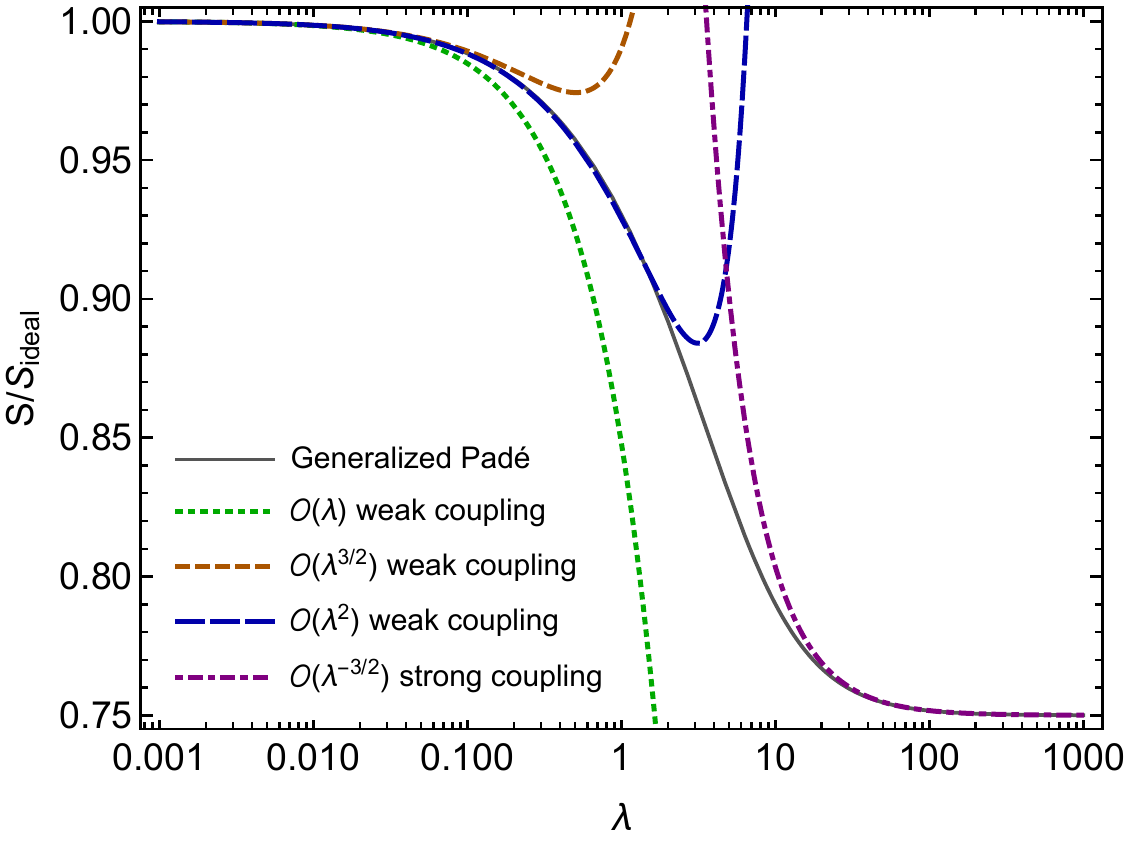}
\end{center}
\vspace{-7mm}
\caption{$\symff$ scaled entropy density ${\cal S}/{\cal S}_\text{ideal}$ as a function of $\lambda$. The green dotted, red dashed, and blue long-dashed curves correspond to the perturbative result truncated at ${\cal O}(\lambda)$, ${\cal O}(\lambda^{3/2})$, and ${\cal O}(\lambda^2)$, respectively.  The purple dot-dashed curve corresponds to the large-$N_c$ strong-coupling result truncated at $O(\lambda^{-3/2})$.  The solid gray line is a generalized Pad\'{e} that interpolates between the known weak- and strong-coupling results.}
\label{fig:finalres}
\end{figure}

In Fig.~\ref{fig:finalres} we plot the scaled entropy density as a function of $\lambda$. The green dotted, red dashed, and blue long-dashed curves correspond to the perturbative result truncated at ${\cal O}(\lambda)$, ${\cal O}(\lambda^{3/2})$, and ${\cal O}(\lambda^2)$, respectively.  The purple dot-dashed curve corresponds to the large-$N_c$ strong-coupling result truncated at $O(\lambda^{-3/2})$.  The solid gray line is a generalized Pad\'{e} that interpolates between the known weak- and strong-coupling results.  For details of the form of the Pad\'{e} constructed and the resulting coefficients, we refer the reader to App.~\ref{app:pade}.  Fig.~\ref{fig:finalres} suggests that the weak-coupling expansion has a rather large radius of convergence in $\symff$ of ${\cal O}(1)$.  One can take the value of $\lambda$ at which the truncated perturbative solutions significantly depart from the Pad\'{e} approximant as an estimate of the range of validity of each perturbative truncation.  From Fig.~\ref{fig:finalres}, when truncated at ${\cal O}(\lambda)$, one must have $\lambda \lesssim 0.02$ in order for the Pad\'{e} approximant and the perturbative result to be approximately equal.  At ${\cal O}(\lambda^{3/2})$, one finds $\lambda \lesssim 0.2$, and at ${\cal O}(\lambda^{2})$, one finds $\lambda \lesssim 2$.  This suggests that adding each perturbative order extends the estimated range of validity in $\lambda$ by an order of magnitude.  In comparison to the convergence of the perturbative QCD free energy we observe that the ${\cal O}(\lambda^2)$ truncation in $\symff$ has ${\cal P}/{\cal P}_\text{ideal} = {\cal S}/{\cal S}_\text{ideal} < 1$ for $\lambda \lesssim 10$, whereas the ${\cal O}(\lambda^2)$ truncation in QCD has ${\cal P} > {\cal P}_{\rm ideal}$ for $\lambda \gtrsim 3.5$ for the central value of the renormalization scale.  In contrast, lattice QCD measurements of the pressure find ${\cal P} < {\cal P}_{\rm ideal}$.  In QCD, one has to include all contributions through ${\cal O}(\lambda^{5/2})$ in order for the pressure to be less than the ideal pressure at large coupling.  This suggests that the perturbative expansion of the $\symff$ free energy might have better convergence than the perturbative expansion of the QCD free energy.

\section{Conclusions and outlook}\la{conclusions}

In this paper we computed the thermodynamic function of $\symff$ to ${\cal O}(\lambda^2)$.  Our final result, presented in eq.~\eqref{eq:finalF}, extends our knowledge of weak-coupling $\symff$ thermodynamics to include terms at ${\cal O}(\lambda^2)$  and ${\cal O}(\lambda^2 \log\lambda)$.  The appearance the ${\cal O}(\lambda^2 \log\lambda)$ contribution can be traced back to non-analytic terms generated due to dressing of the gluon and scalar propagators at finite temperature (ring resummation).  All results presented here made use of the RDR scheme which manifestly preserves supersymmetry.  Having obtained the ${\cal O}(\lambda^2)$  and ${\cal O}(\lambda^2 \log\lambda)$ coefficients in the $\symff$ free energy, we then constructed a large-$N_c$ Pad\'{e} approximant that interpolates between the weak- and strong-coupling limits.  The resulting singularity-free Pad\'{e} approximant \eqref{eq:pade} reproduces the weak-coupling limit through ${\cal O}(\lambda^2,\lambda^2 \log\lambda)$ and the coefficients and analytic structure of the large-$N_c$ strong-coupling limit through ${\cal O}(\lambda^{-5/2})$, with no terms containing $\log\lambda$ appearing through this order.

In the near future we plan to also compute the coefficient of $\lambda^{5/2}$ in the $\symff$ free energy. For this purpose, one can either adapt the methods presented originally by Zhai and Kastening in QCD~\cite{Zhai:1995ac}  to $\symff$ or one could consider applying effective field theory methods similar to refs.~\cite{Braaten:1995jr} and \cite{Nieto:1999kc}.  We plan to pursue the first option in the near term due the fact that the results presented herein followed the Arnold, Zhai, and Kastening calculational framework.  It would certainly be interesting to consider the application of effective field theory methods to $\symff$ thermodynamics since, as emphasized herein, the necessary three-loop diagrams can be evaluated using dimensional reduction of the simpler $\symot$ theory.  Finally, we mention that we also plan to pursue a three-loop HTLpt calculation of $\symff$ thermodynamics in order to further improve the convergence of the successive approximations to the thermodynamic functions in the weak-coupling limit.  This would extend our prior two-loop HTLpt calculation of $\symff$ thermodynamics to three-loop order~\cite{Du:2020odw}.

The future work outlined above will extend the perturbative calculation of the $\symff$ free energy to the highest order possible before genuinely non-perturbative effects related to magnetic scales in SYM enter \cite{Nieto:1999kc}.  Like QCD, in $\symff$ at ${\cal O}({\lambda^3})$ there will be both perturbative and non-perturbative contributions.  The non-perturbative contribution at ${\cal O}({\lambda^3})$ can be computed on the lattice using effective field theory methods, however, computation of the four-loop perturbative contributions at this order presents a technical challenge.

\section*{Note added}
In arXiv version 3, we corrected an error in eq.~\eqref{f3loopn4counter2}, changing $(D-2) \rightarrow (d-2)$ plus some associated clarifiations in eqs.~\eqref{shadedblobg}-\eqref{diag4eq}.  The change to eq.~\eqref{f3loopn4counter2} resulted in modifications of eq.~\eqref{3loopresum}, eq.~\eqref{eq:finalF}, fig.~\ref{fig:finalres}, and the expression for $b$ in eq.~\eqref{padecoeffs} compared to version 2.  These changes has been verified by computing the free energy using effective field theory methods.  Importantly, our conclusions are unchanged since the correction is numerically small, however, an erratum correcting this error has been submitted to JHEP.  We have also taken this opportunity to correct some typos.

\section*{Acknowledgements}
We thank J.O. Andersen for calling the issue mentioned above to our attention.  We thank S. Grozdanov, J. Maldacena, S.  Martin,  A. Rebhan, P. Romatschke, W. Seigel, A. Starinets, M. A. Vazquez-Mozo, and V. N. Velizhanin for helpful discussions.  Q.D., M.S., and U.T. were supported by the U.S. Department of Energy, Office of Science, Office of Nuclear Physics under Award No.~DE-SC0013470.  In addition, Q.D. was supported by the China Scholarship Council under Project No.~201906770021, the National Natural Science Foundation of China Project No.~11935007, and the Guangdong Major Project of Basic and Applied Basic Research No. 2020B0301030008.

\appendix

\section{Feynman rules for resummed $\symff$ }\la{propagatpr}

Based on the Lagrangian density in eq.~\eqref{lagresum} one sees that only the gluon and scalar propagator are modified using this resummation method and, as a consequence, there will be two thermal mass counterterm vertices which must be taken into account.

\subsection{The resummed gluon propagator}\la{gpropagatpr}

The Feynman rule for the resummed gluon propagator is
\ba\la{prog}
-i \delta^{ab}\Delta_{\mu\nu}(p) \,,
\ea
where the tensor-valued gluon propagator $\Delta_{\mu\nu}$ depends on the choice of gauge fixing.  In covariant gauge it can be expressed as
\ba\la{prog1}
\Delta_{\mu\nu}(p)=\frac{g^{\mu\nu}}{p^2} -(1-\xi)\frac{p^\mu p^\nu}{p^4}+\bigg(\frac{1}{p^2-m_D^2}-\frac{1}{p^2}\bigg)\delta_{p_0} g^{\mu 0}g^{\nu 0}\,.
\ea
%

\subsection{The resummed scalar propagator}\la{spropagatpr}
The Feynman rule for the resummed scalar propagator is
\ba\la{pros}
i \delta^{ab}\delta^{AB}\Delta_s(p) \,,
\ea
where
\ba\la{pros1}
\Delta_s(p)=\frac{1}{p^2} +\bigg(\frac{1}{p^2-M^2}-\frac{1}{p^2}\bigg)\delta_{p_0} \, .
\ea
Note that this is the same form as the $\lambda\phi^4$ resummed propagator presented in ref.~\cite{Andersen:2004fp}.

\subsection{The counterterm vertex}\la{ctvertex}

The gluonic counterterm vertex is
\ba\la{gcount}
-i \delta^{ab} m_D^2 \delta_{p_0} g^{\mu 0} g^{\nu 0}\,.
\ea
The scalar counterterm vertex is
\ba\la{scount}
i \delta^{ab} \delta^{AB} M^2 \delta_{p_0} \,.
\ea
%

\section{Feynman rules for $\symod$ }\la{propagatorn1}

From the Lagrangian density in eq.~(\ref{actN1}), one finds that the gluon and ghost propagators, the three/four gluon vertex, and the gluon-ghost vertex are all the same as in QCD, however, the dimension of the momentum-space and the metric tensor changes from $4$ to $\mathcal{D}$.

\subsection{The quark propagator}\la{qpropagatprn1}

The Feynman rule for the quark propagator is
\ba\la{quark}
 \frac{i \delta^{ab}}{p\cdot \Gamma}\,.
\ea
%

\subsection{The quark-gluon vertex}\la{qgvertex}
The Feynman rule for the quark-gluon vertex is
\ba\la{qgvertexeq}
gf^{abc}\Gamma^M\,.
\ea

\section{One-loop self-energies in $\symod$}\la{self_energyn1}

\begin{figure}[t]
\begin{center}
\includegraphics[width=1\linewidth]{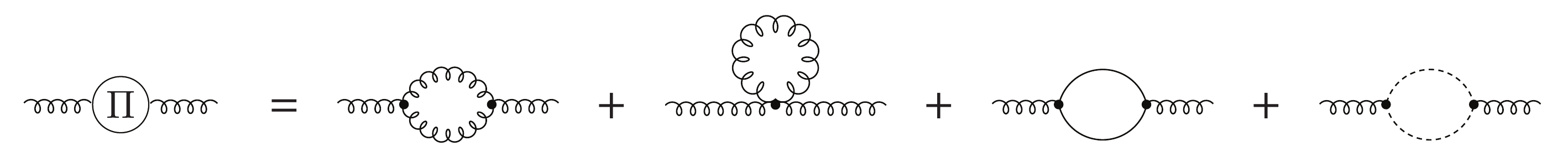}
\end{center}
\vspace{-4mm}
\caption{The one-loop self-energy of $\mathcal{N}=1$ SYM.}
\label{fig:diagrams5}
\end{figure}

In the following, for convenience, we use ${{\Large\int}\hspace{-0.27cm}{\tiny\Sigma}}_{Q,d}$ to represent the Euclidean-space momentum integral in $\symod$. The one-loop self-energy of the gauge bosons $(\Pi_{MN}^{ab} (P))_{\symod}$ contains two parts
\ba
\Pi^{b,ab}_{MN}(P) &=& -\lambda \delta^{ab} \bigg[ \frac{{\mathcal D}-2}{2} \bar{\Pi}^b_{MN}(P) -2 (P^2\delta_{MN}-P_M P_N)\sumint_{Q,d}  \frac{1}{Q^2(P+Q)^2}  \bigg], \la{self_energy_n11} \;\; \\
\Pi^{f,ab}_{MN}(P) &=& \lambda \delta^{ab} \frac{\textrm{Tr}\,I_n}{2} \bigg[ \bar{\Pi}^f_{MN}(P) - (P^2\delta_{MN}-P_M P_N)\sumint_{\{Q,\},d} \frac{1}{Q^2(P+Q)^2}  \bigg], \la{self_energy_n12}
\ea
where
\ba
\bar{\Pi}^b_{MN}(P) &\equiv& 2\delta_{MN}\sumint_{Q,d} \frac{1}{Q^2}-\sumint_{Q,d} \frac{(2Q+P)_M (2Q+P)_N}{Q^2(P+Q)^2}, \la{self_energy_n13} \\
\bar{\Pi}^f_{MN}(P) &\equiv& 2\delta_{MN}\sumint_{\{Q\},d} \frac{1}{Q^2}-\sumint_{\{Q\},d} \frac{(2Q+P)_M (2Q+P)_N}{Q^2(P+Q)^2} \, , \la{self_energy_n14}
\ea
and we have taken ${\cal D} = {\cal D}_{\rm max}$.  One can obtain $(\Pi_{MN}^{ab} (P))_{\symod}$ directly from the corresponding result in QCD. From the Lagrangian, in the bosonic sector the only difference between $\symod$ and QCD is the dimension of the metric tensor and momenta.  As a result, one can immediately generalize the one-loop self-energy results from QCD.  In the fermionic contributions an additional difference stems from the different fermionic spinor and color representations in the two theories, which causes $\textrm{Tr} \, I_n=\mathcal{D}-2$ and $S_F=N_c$.\footnote{$S_F \equiv \textrm{Tr}\,T^aT^a/d_A$, with $T^a$ the generator for the fermionic color algebra. $S_F=N_f/2$ in the fundamental representation with $N_f$ being the number of quark flavors, while $S_F=N_c$ in the adjoint representation.}

In the hard-thermal-loop limit (HTL), by using the contour integral method and integration by parts, one obtains the bosonic thermal mass in $\symod$
\ba\la{mDn1}
(m_D^2)_d^{\mathcal{D}} = \lambda ( \mathcal{D}-2)(d-2) \bigg[ \sumint_{Q,d} \frac{1}{Q^2} - \sumint_{\{Q\},d} \frac{1}{Q^2}  \bigg] ,
\ea
where $d={\cal D}$.  We note that eq.~(\ref{mDn1}) is the same as the result that would be obtained by using $-\delta^{MN} \Pi_{MN}$, but for ease of application of the RDR scheme, we choose the form in eq.~(\ref{mDn1}).

In the same way, by calculating the one-loop fermionic self-energy in the HTL limit, one obtains the quark thermal mass
\ba\la{mqn1}
(m_q^2)_d^{\mathcal{D}} = \frac{\lambda}{2} ( \mathcal{D}-2) \bigg[ \sumint_{Q,d} \frac{1}{Q^2} - \sumint_{\{Q\},d} \frac{1}{Q^2}  \bigg].
\ea

By imposing $\mathcal{D}_{\textrm{max}}=10$ and $d=4-2\epsilon$ in (\ref{mDn1}) and (\ref{mqn1}), one obtains
\ba
(m_D^2)_{4-2\epsilon}^{10} & = & 8\lambda (d-2) [b_1 - f_1]=2 \lambda T^2 + \mathcal{O}(\epsilon)\, , \la{mDmassnew} \\
(m_q^2)_{4-2\epsilon}^{10} & = & 4 \lambda [ b_1 - f_1 ] = \frac{1}{2} \lambda T^2 + \mathcal{O}(\epsilon)\, .\la{mqmassnew}
\ea
Note that these are the same as the bosonic and fermionic thermal masses with $D=4$ in $\symff$.  However, in the $\symot$ theory one cannot obtain the thermal mass for scalars straightforwardly since they do not appear as fundamental degrees of freedom.  As a consequence, one cannot use the RDR method based on massive $\symot$ diagrams and one must perform the necessary resummations directly in $\symff$.

\section{Derivations of $I_{\symod}^{\textrm{bb}}$, $I_{\symod}^{\textrm{bf}}$, and $I_{\symod}^{\textrm{ff}}
$ }\la{In1bf}

The integrals $I_{\symod}^{\textrm{bb}}$, $I_{\symod}^{\textrm{bf}}$, and $I_{\symod}^{\textrm{ff}}$ are defined by integration of the $\symod$ one-loop self-energies presented in App.~\ref{self_energyn1}
\ba
\lambda^2 I_{\symod}^{\textrm{bb}} &=& \sumint_{P,d} \frac{ [\Pi^b_ {MN}(P) ]^2 }{P^4} \, ,  \la{In1symbb1}\\
\lambda^2 I_{\symod}^{\textrm{ff}} &=& \sumint_{P,d} \frac{[\Pi^f_ {MN} (P)]^2 }{P^4} \, ,  \la{In1symff1}\\
\lambda^2 I_{\symod}^{\textrm{bf}} &=& \sumint_{P,d} \frac{2\Pi^b_ {MN} (P) \Pi^f_ {MN} (P) }{P^4} \, ,  \la{In1symbf1}
\ea
where ${\cal D} = {\cal D}_\text{max}$.
Inserting (\ref{self_energy_n11}) and (\ref{self_energy_n12}) and expanding, one obtains three terms we define as
\ba
\bar{I}^{\textrm{bb}}_{\symod} &\equiv& \sumint_{P,4-2\epsilon} \frac{ [\bar{\Pi}^b_ {MN}(P) ]^2 }{P^4}  = 4({\mathcal D}-4)b_2 b_1^2 +16 H_5-I_{\textrm{ball}}^{\textrm{bb}} \, ,  \la{In1symbb2}\\
\bar{I}^{\textrm{ff}}_{\symod} &\equiv& \sumint_{P,4-2\epsilon} \frac{[\bar{\Pi}^f_ {MN} (P)]^2 }{P^4} = 4({\mathcal D}-4)b_2 f_1^2 +16 H_4-I_{\textrm{ball}}^{\textrm{ff}}  \, ,  \la{In1symff2}\\
\bar{I}^{\textrm{bf}}_{\symod} &\equiv& \sumint_{P,4-2\epsilon} \frac{\bar{\Pi}^b_ {MN} (P) \bar{\Pi}^f_ {MN} (P) }{P^4} =  4({\mathcal D}-4)b_2 b_1 f_1 +16 H_6 - I_{\textrm{ball}}^{\textrm{bf}} \, ,  \la{In1symbf2}
\ea
where $H_4$, $H_5$ and $H_6$ are defined by
\ba
H_4 & \equiv &  \sumint_{P\{QK\}}  \frac{(Q\cdot K)^2}{P^4 Q^2 K^2(P+Q)^2 (P+K)^2} \, , \la{Hfun2} \\
H_5 & \equiv &  \sumint_{PQK} \frac{(Q\cdot K)^2}{P^4 Q^2 K^2(P+Q)^2 (P+K)^2} \, , \la{Hfun3}  \\
H_6 & \equiv &  \sumint_{PQ\{K\}} \frac{(Q\cdot K)^2}{P^4 Q^2 K^2(P+Q)^2 (P+K)^2} \la{Hfun4} \, .
\ea
The final results for these integrals are given in app.~\ref{HfunAppen}. Simplifying and using
\ba\la{Isum}
\sumint_{PQ/P\{Q\}/\{PQ\}} \frac{1}{P^2Q^2(P+Q)^2}=0 \, ,
\ea
we obtain the final results given in eqs.~(\ref{In1symbb1}), (\ref{In1symff1}), and (\ref{In1symbf1}) in $d=4-2\epsilon$ dimensions
\ba
I_{\symod}^{\textrm{bb}}\big|_{d=4-2\epsilon} &=& \frac{({\mathcal D}-2)^2}{4} \bar{I}^{\textrm{bb}}_{\symod} + 2 {\mathcal D} I_{\textrm{ball}}^{\textrm{bb}} \, ,  \la{In1symbb3} \\
I_{\symod}^{\textrm{ff}} \big|_{d=4-2\epsilon} &=& \frac{(\textrm{Tr} \, I_n)^2}{4} \big[ \bar{I}^{\textrm{ff}}_{\symod} +({\mathcal D}-3)I_{\textrm{ball}}^{\textrm{ff}} \big] ,  \la{In1symff3}\\
I_{\symod}^{\textrm{bf}} \big|_{d=4-2\epsilon} &=& -\textrm{Tr} \, I_n \bigg[ \frac{{\mathcal D} -2}{2}\bar{I}^{\textrm{bf}}_{\symod} + \frac{3}{2}({\mathcal D}-2)I_{\textrm{ball}}^{\textrm{bf}}  \bigg] .  \la{In1symbf3}
\ea

\section{Derivations of $H_4$, $H_5$, and $H_6$ in $d=(4-2\epsilon)$-dimensions}\la{HfunAppen}

The simplest way to calculate $H_4$, $H_5$, and $H_6$ in eqs.~(\ref{Hfun2}), (\ref{Hfun3}), and (\ref{Hfun4}) is to use the method introduced in ref.~\cite{Arnold:1994eb}, which is to use the final results for $I_{\textrm{sqed}}^{\textrm{bb}}$, $I_{\textrm{sqed}}^{\textrm{bf}}$, and $I_{\textrm{sqed}}^{\textrm{ff}}$, defined by
\ba
I_{\textrm{sqed}}^{\textrm{bb}} &\equiv& \sumint_{P} \frac{ \big[\Delta \bar{\Pi}_{\mu\nu}^{b}(P) \big]^2 }{P^4},  \la{Isqedbb}\\
I_{\textrm{sqed}}^{\textrm{bf}} &\equiv& \sumint_{P} \frac{ \big[\Delta \bar{\Pi}_{\mu\nu}^{b}(P) \Delta \bar{\Pi}_{\mu\nu}^{f}(P) \big]  }{P^4},    \la{Isqedbf}\\
I_{\textrm{sqed}}^{\textrm{ff}} &\equiv& \sumint_{P} \frac{ \big[\Delta \bar{\Pi}_{\mu\nu}^{f}(P) \big]^2 }{P^4},  \la{Isqedff}
\ea
where $\Delta\bar{\Pi}_{\mu\nu}(P) \equiv \bar{\Pi}_{\mu\nu}(P)-\bar{\Pi}^{\rho\rho}(0) \delta_{\mu0} \delta_{\nu0} \delta_{p_0}$.
For the calculation of $H_4$, using eq.~(\ref{self_energy_n4b4}) in eq.~(\ref{Isqedff}) and simplifying, one obtains
\ba\la{Isqedfffinal}
I_{\textrm{sqed}}^{\textrm{ff}} = 4(d-4)b_2 f_1^2 + 16H_4 -I_{\textrm{ball}}^{\textrm{ff}} \, .
\ea
Since $I_{\textrm{sqed}}^{\textrm{ff}}$ has been calculated previously, with the result being
\ba\la{Isqedfffnumerical}
I_{\textrm{sqed}}^{\textrm{ff}} = \frac{1}{(4\pi)^2}\bigg(\frac{T^2}{12} \bigg)^2 \bigg[\frac{11}{6\epsilon} & +& 11\log\frac{\bar{\mu}}{4\pi T}+\frac{1}{3} \frac{\zeta'(-3)}{\zeta(-3)}+\frac{20}{3}\frac{\zeta'(-1)}{\zeta(-1)} +4 \gamma_E   \nonumber \\& + & \frac{281}{60} -13 \log 2\bigg]+\mathcal{O}(\epsilon) \, ,
\ea
one obtains the following result for $H_4$
\ba\la{H4}
H_4 = \frac{1}{(4\pi)^2}\bigg(\frac{T^2}{12} \bigg)^2 \bigg[\frac{5}{24\epsilon} & +& \frac{5}{4}\log\frac{\bar{\mu}}{4\pi T} - \frac{1}{6} \frac{\zeta'(-3)}{\zeta(-3)}+\frac{7}{6}\frac{\zeta'(-1)}{\zeta(-1)} + \frac{1}{4} \gamma_E   \nonumber \\& + & \frac{23}{24} -\frac{8}{5} \log 2\bigg]+\mathcal{O}(\epsilon) \,.
\ea

From eq.~(\ref{Isqedfffinal}) one finds $I_{\textrm{sqed}}^{\textrm{ff}}$ is independent of the thermal mass contribution because $\textrm{I}_{\textrm{resum}}^f=0$. This will not occur in $I_{\textrm{sqed}}^{\textrm{bb}}$ and $I_{\textrm{sqed}}^{\textrm{bf}}$ since $\textrm{I}_{\textrm{resum}}^b\neq 0$.  By using eqs.~(\ref{Isqedbb}), (\ref{Iballbb}), (\ref{Isqedbf}), and (\ref{Iballbf}), the final results for $H_5$ and $H_6$ can be obtained
\ba
H_5 &=& \frac{1}{(4\pi)^2}\bigg(\frac{T^2}{12} \bigg)^2 \bigg[\frac{4}{3\epsilon} + 8\log\frac{\bar{\mu}}{4\pi T} - \frac{5}{3} \frac{\zeta'(-3)}{\zeta(-3)}+\frac{26}{3}\frac{\zeta'(-1)}{\zeta(-1)} +\gamma_E + \frac{49}{12} \bigg] +\mathcal{O}(\epsilon) \, ,   \la{H5} \;\; \\
H_6 &=& \frac{1}{(4\pi)^2}\bigg(\frac{T^2}{12} \bigg)^2 \bigg[-\frac{17}{48\epsilon} - \frac{17}{8}\log\frac{\bar{\mu}}{4\pi T} + \frac{5}{24} \frac{\zeta'(-3)}{\zeta(-3)}-\frac{11}{6}\frac{\zeta'(-1)}{\zeta(-1)} -\frac{1}{2}\gamma_E - \frac{41}{48} \nonumber \\&& \qquad\qquad\qquad\qquad\quad +\frac{11}{8}\log2 \bigg] +\mathcal{O}(\epsilon) \, .   \la{H6}
\ea

\section{One-loop bosonic self-energies in $\symff$}\la{self_energyn4}

\begin{figure}[t]
\begin{center}
\includegraphics[width=0.9\linewidth]{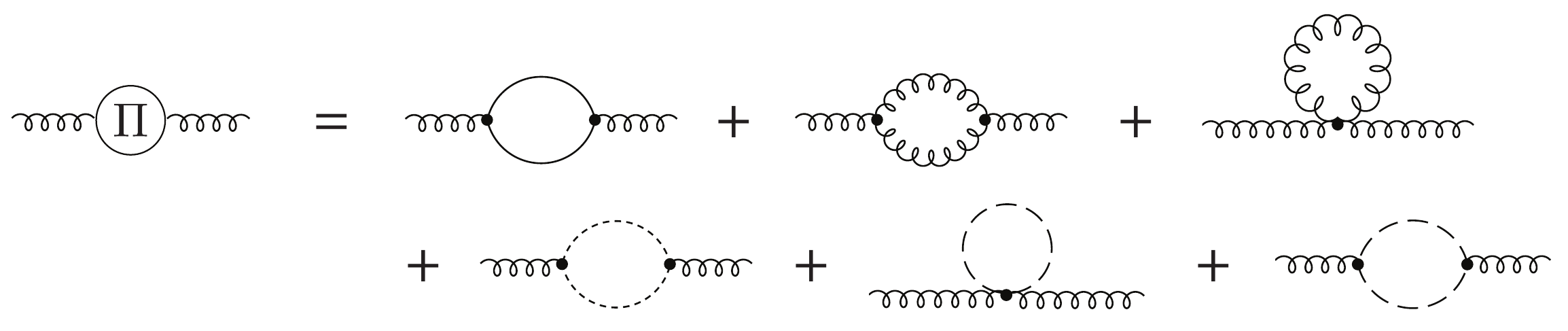}
\end{center}
\vspace{-4mm}
\caption{The gluon self-energy in $\symff$.}
\label{fig:diagrams6}
\end{figure}

The one-loop gauge bosonic self-energy of $(\Pi_{\mu\nu}^{ab} (P))_{\symff}$ shown in Fig.~\ref{fig:diagrams6} contains two parts,
\ba
\Pi^{b, ab}_{\mu\nu}  (P) &=& -\lambda \delta^{ab} \bigg[ \frac{D+4}{2} \bar{\Pi}^b_{\mu\nu}(P) -2 (P^2 \delta_{\mu\nu}-P_\mu P_\nu)\sumint_Q \frac{1}{Q^2(P+Q)^2}  \bigg], \la{self_energy_n4b1} \\
\Pi^{f, ab}_{\mu\nu} (P) &=& 2\lambda \delta^{ab} \bigg[ 2 \bar{\Pi}^f_{\mu\nu}(P) -2 (P^2 \delta_{\mu\nu}-P_\mu P_\nu)\sumint_{\{Q\}} \frac{1}{Q^2(P+Q)^2}  \bigg], \la{self_energy_n4b2}
\ea
where
\ba
\bar{\Pi}^b_{\mu\nu}(P) &\equiv& 2\delta_{\mu\nu}\sumint_Q \frac{1}{Q^2}-\sumint_Q \frac{(2Q+P)_\mu (2Q+P)_\nu}{Q^2(P+Q)^2}, \la{self_energy_n4b3} \\
\bar{\Pi}^f_{\mu\nu}(P) &\equiv& 2\delta_{\mu\nu}\sumint_{\{Q\}} \frac{1}{Q^2}-\sumint_{\{Q\}} \frac{(2Q+P)_\mu (2Q+P)_\nu}{Q^2(P+Q)^2}. \la{self_energy_n4b4}
\ea
One finds that the form in eq.~(\ref{self_energy_n4b1}) is the same as QCD except for the coefficient of $\bar{\Pi}^b_{\mu\nu}(P)$, which is due to the fact that there are 6 extra scalars in $\symff$ compared to QCD. The difference in $\Pi^{f, ab}_{\mu\nu} (P)$ occurs because there are 4 Majorana fermions in the adjoint representation which are their own antiparticles.

\begin{figure}[t]
\vspace{6mm}
\begin{center}
\includegraphics[width=0.975\linewidth]{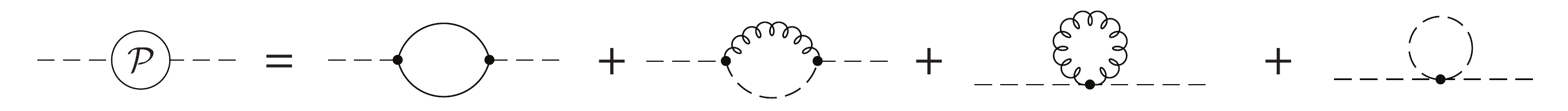}
\end{center}
\vspace{-4mm}
\caption{The scalar self-energy in $\symff$.}
\label{fig:diagrams7}
\end{figure}

The one-loop scalar self-energy in $\symff$ can be expressed as
\ba
\mathcal{P}^{ab,AB}(P) &\!=\!& \lambda \delta^{ab} \delta^{AB} \bigg\{ \sumint_Q \bigg[ \frac{(D+4)}{Q^2}- \frac{2P^2}{Q^2(P+Q)^2} \bigg] \! 
- \sumint_{\{Q\}} \bigg[ \frac{8}{Q^2} - \frac{4P^2}{Q^2(P+Q)^2}  \bigg] \bigg\} , \hspace{8mm}
\ea
where we present the individual diagrams contributing in fig.~\ref{fig:diagrams7}. The scalar thermal mass is 
\be
M^2= \mathcal{P}(0) = \lambda [(D+4)b_1-8f_1]=\lambda T^2 + \mathcal{O}(\epsilon) \, .\la{Mmassnew}
\ee

\section{Large-$N_c$ generalized Pad\'{e} approximant}
\label{app:pade}

With new perturbative coefficients in hand one can produce an updated Pad\'{e} approximant along the lines presented in refs.~\cite{Kim:1999sg,Blaizot:2006tk}.  For this purpose one can make use of the fact that, in the large-$N_c$ limit, the strong-coupling expansion becomes a series in powers of $\lambda^{-3/2}$ only.  Using this information, one can construct a constrained large-$N_c$ interpolating function that connects the weak coupling expansion and the strong-coupling expansion.  We note, however, that if one considers sub-leading corrections in $1/N_c$ in the strong-coupling limit, the entropy density can contain additional fractional powers of $\lambda$ \cite{Myers:2008yi} and, in this case, there may also be logarithms induced by massless gravity modes.  Since knowledge of such corrections is limited, we focus on constructing a Pad\'{e} approximant that is valid in the large-$N_c$ limit only.\footnote{Exact solutions for the entropy density in 2+1D $O(N)$ conformal field theories in the large-$N$ limit have a similar analytic structure, namely logarithms of the coupling appearing in the weak-coupling limit, but not in the strong-coupling limit \cite{Romatschke:2019ybu}.} In order for a Pad\'{e} approximant to approach a constant in the strong-coupling limit it must be a symmetric or balanced Pad\'{e} approximant, in which the maximal powers of $\lambda$ appearing in the numerator and denominator are the same.

Based on the large-$N_c$ structure of the strong-coupling expansion, we find that the following form can reconstruct all known coefficients in both the weak- and strong-coupling limits
\be
\frac{S}{S_{\rm ideal} } =  \frac{1 + a \lambda^{1/2}  + b \lambda  + c \lambda^{3/2}  + d \lambda^2  + e \lambda^{5/2}  }{1 + a \lambda^{1/2}  + \bar{b} \lambda  + \frac{4}{3} c \lambda^{3/2}  + \frac{4}{3} d \lambda^2  + \frac{4}{3} e \lambda^{5/2}  }  \; ,
\label{eq:pade}
\ee
with
{\footnotesize
\ba
a &=& \frac{4 \pi ^2}{135 \zeta (3)}+\frac{2 \left(3+\sqrt{2}\right)}{3 \pi } \, , \nonumber \\
b &=& \frac{1}{\pi^2} \log\left( \frac{\lambda}{\pi^2} \right) +\frac{16 \pi  \left[ 45 \left(3+\sqrt{2}\right) \zeta (3)+\pi
   ^3\right]}{18225 \zeta^2(3)}+ \frac{36 \left[ \frac{\zeta'(-1)}{\zeta(-1)}+\gamma_E \right] +69 \sqrt{2}+59-75 \log 2}{36 \pi ^2} \, , \nonumber  \\
&& \nonumber  \\
\bar{b} &=& b + \frac{3}{2 \pi ^2} \, ,  \nonumber  \\
c &=& \frac{2}{15 \zeta (3)} \, , \nonumber  \\
d &=& \frac{180 \left(3+\sqrt{2}\right) \zeta (3)+8 \pi ^3}{2025 \pi  \zeta^2(3)} \, , \nonumber  \\
e &=& \frac{2b}{15 \zeta (3)}-\frac{3}{5 \pi ^2 \zeta (3)} \, . \la{padecoeffs}
\ea
}
The coefficients of 4/3 in the denominator of eq.~\eqref{eq:pade} ensure that in the strong-coupling limit (a) one obtains the correct asymptotic limit of 3/4 and that (b) terms of the form $\lambda^{-1/2}$, $\lambda^{-1/2} \log\lambda$, $\lambda^{-1}$, and $\lambda^{-1} \log\lambda$ do not appear in the series expansion.  To fix the remaining coefficients in eq.~\eqref{eq:pade} one uses the explicit expressions for the strong and weak coupling expansions through ${\cal O}(\lambda^{-3/2})$ and  ${\cal O}(\lambda^2,\lambda^2\log\lambda)$ provided by eqs.~\eqref{eq:sclimit} and \eqref{eq:finalF}, respectively.  In the weak-coupling limit, this form reproduces the perturbative result through ${\cal O}(\lambda^2,\lambda^2\log\lambda)$.  In the strong-coupling limit, \eqref{eq:pade} reproduces the known result through ${\cal O}(\lambda^{-3/2})$, with the next non-vanishing strong-coupling contribution appearing at ${\cal O}(\lambda^{-3})$, i.e. there are no contributions occurring at order ${\cal O}(\lambda^{-2})$, ${\cal O}( \lambda^{-2} \log\lambda)$, ${\cal O}(\lambda^{-5/2})$, and ${\cal O}( \lambda^{-5/2} \log\lambda)$.

We note that, at the order used, namely through ${\cal O}(\lambda^{5/2})$ in the numerator and denominator, the coefficients in the Pad\'{e} approximant \eqref{eq:pade} obtained are unique.  That said, it is possible to add additional terms in both the numerator and denominator, e.g. extending to including ${\cal O}(\lambda^3)$ terms, which would introduce new coefficients that are unconstrained due to limited information from the weak- and strong-coupling expansions.  Based on current information, our final result \eqref{eq:pade} represents the highest order Pad\'{e} approximant for which all coefficients can be uniquely constrained.

\bibliographystyle{JHEP}
\bibliography{susy3}

\end{document}